\documentclass[10pt]{article}

\usepackage{amsmath}
\usepackage{array}
\usepackage{appendix}
\usepackage{graphicx}
\usepackage{amsfonts}
\usepackage{amssymb}
\usepackage{mathrsfs}
\usepackage{yfonts}
\usepackage{euscript}
\usepackage{upgreek}
\usepackage{slantsc}
\usepackage{calligra}
\usepackage[T1]{fontenc}
\usepackage{epsf}
\usepackage{latexsym}

\usepackage{tipa}

\def\be{\begin{equation}}
\def\ee{\end{equation}}
\def\beq{\begin{equation}}
\def\eeq{\end{equation}}
\def\bea{\begin{eqnarray}}
\def\eea{\end{eqnarray}}

\def\ni{\noindent}

\def\!{\hspace{-1.6667em}}

\def\mA{\mbox{A}}   
  
\def\mC{\mbox{C}}   
\def\mD{\mbox{D}}

\def\mF{\mbox{F}}
\def\mG{\mbox{G}}
 
\def\mI{\mbox{I}}
\def\mJ{\mbox{J}}  
\def\mK{\mbox{K}}
\def\mL{\mbox{L}}
\def\mM{\mbox{M}}
\def\mN{\mbox{N}} 

\def\mP{\mbox{P}}
\def\mQ{\mbox{Q}}
\def\mR{\mbox{R}}
\def\mS{\mbox{S}}
\def\mT{\mbox{T}} 

\def\mV{\mbox{V}}
\def\mW{\mbox{W}}
\def\mX{\mbox{X}}
\def\mY{\mbox{Y}}
\def\mZ{\mbox{Z}} 

\def\mg{\mbox{g}}
\def\mh{\mbox{h}}

\def\mn{\mbox{n}}   

\def\mp{\mbox{p}}

\def\ms{\mbox{s}}
\def\mt{\mbox{t}}
\def\muu{\mbox{u}}
\def\mv{\mbox{v}} 
\def\mw{\mbox{w}}

\def\slLambda{\mathit{\Lambda}}               

\def\uupbeta{{\underline{\upbeta}}}

\def\brho{\mbox{\boldmath$\rho$}}          
\def\uupbeta{\mbox{\underline{$\upbeta$}}}

\def\bupSigma{\mbox{\boldmath$\Sigma$}}                 
\def\bupPhi{\mbox{\boldmath$\Phi$}}                     
\def\sbSigma{\mbox{\scriptsize\boldmath$\Sigma$}}

\def\cH{{\mathscr H}}

\def\cJ{{\mathscr J}}
\def\cK{{\mathscr K}}
\def\cL{{\mathscr L}}

\def\cS{{\mathscr S}}
\def\cT{{\mathscr T}}

\def\cs{\mbox{\scriptsize$\cS$}}

\def\ux{\underline{{x}}}

\def\bh{\underline{\underline{\mbox{h}}}  }            

\def\suF{\underline{\mbox{\scriptsize F}}}

\def\bp{\mbox{\bf p}}

\def\bK{\mbox{\bf K}}

\def\bM{\mbox{\bf M}}

\def\bg{\mbox{{\bf g}}}
\def\bM{\mbox{{\bf M}}}
\def\bM{\mbox{{\bf M}}}
\def\bN{\mbox{{\bf N}}}

\def\bh{\mbox{{\bf h}}}
\def\bm{\mbox{{\bf m}}}

\def\scC{\mbox{\scriptsize ${\cal C}$}}          
\def\scE{\mbox{\scriptsize ${\cal E}$}}          
\def\scF{\mbox{\scriptsize ${\cal F}$}}
\def\scH{\mbox{\scriptsize ${\cal H}$}}          
\def\scI{\mbox{\scriptsize ${\cal I}$}}

\def\scL{\mbox{\scriptsize ${\cal L}$}}          
\def\scM{\mbox{\scriptsize ${\cal M}$}}          
\def\scN{\mbox{\scriptsize ${\cal N}$}}
\def\scO{\mbox{\scriptsize ${\cal O}$}}
\def\scP{\mbox{\scriptsize ${\cal P}$}}
\def\scR{\mbox{\scriptsize ${\cal R}$}}          
\def\scS{\mbox{\scriptsize ${\cal S}$}}

\def\scU{\mbox{\scriptsize ${\cal U}$}}          

\def\iB{\mbox{\scriptsize$B$}}   
\def\iD{\mbox{\scriptsize$D$}}   
\def\iK{\mbox{\scriptsize$K$}}   

\def\FrQ{\mbox{\Large $\mathfrak{q}$}}
\def\FrT{\mathfrak{T}}                            
\def\FrN{\mathfrak{N}}                            

\def\FrU{\mbox{\boldmath$\mathfrak{U}$}}
\def\FrW{\mbox{\boldmath$\mathfrak{W}$}}          

\def\FrT{\mbox{\boldmath$\mathfrak{T}$}}                        

\def\FrF{\mbox{$\mathfrak{F}$}}                                 
\def\FrL{\mbox{$\mathfrak{L}$}}                                 
 
\def\FrM{\mbox{\Large $\mathfrak{m}$}}                         
\def\FrI{\mbox{$\mathfrak{I}$}}                                 
\def\FrMgen{\mbox{\boldmath$\mathfrak{M}$}}                     
\def\FrHgen{\mbox{\boldmath$\mathfrak{H}$}}                     

\def\sFG{\mbox{$\mathfrak{g}$}}

\def\FrS{\mbox{\Large $\mathfrak{s}$}}

\def\FrG{\mbox{\Large $\mathfrak{g}$}}  
 
\def\FS{\mbox{\LARGE\tt s}}                         

\def\sa{\mbox{\scriptsize a}}

\def\scc{\mbox{\scriptsize c}}

\def\se{\mbox{\scriptsize e}}
\def\sf{\mbox{\scriptsize f}}
\def\sg{\mbox{\scriptsize g}} 
 
\def\si{\mbox{\scriptsize i}}

\def\sll{\mbox{\scriptsize l}}  
\def\sm{\mbox{\scriptsize m}}
\def\sn{\mbox{\scriptsize n}} 
 
\def\sp{\mbox{\scriptsize p}}

\def\sr{\mbox{\scriptsize r}}
\def\st{\mbox{\scriptsize t}}

\def\sA{\mbox{\scriptsize A}} 
\def\sB{\mbox{\scriptsize B}}

\def\sD{\mbox{\scriptsize D}}

\def\sF{\mbox{\scriptsize F}}

\def\sJ{\mbox{\scriptsize J}}
\def\sK{\mbox{\scriptsize K}}
\def\sL{\mbox{\scriptsize L}} 
\def\sM{\mbox{\scriptsize M}}

\def\sP{\mbox{\scriptsize P}} 
 
\def\sR{\mbox{\scriptsize R}}
\def\sS{\mbox{\scriptsize S}}

\def\sW{\mbox{\scriptsize W}}
\def\sX{\mbox{\scriptsize X}}

\def\sfA{\mbox{\sffamily{\scriptsize A}}}      
\def\sfB{\mbox{\sffamily{\scriptsize B}}}      
\def\sfC{\mbox{\sffamily{\scriptsize C}}}      
\def\sfD{\mbox{\sffamily{\scriptsize D}}}      
\def\sfG{\mbox{\sffamily{\scriptsize G}}}      
\def\sfI{\mbox{\sffamily{\scriptsize I}}}      
\def\sfK{\mbox{\sffamily{\scriptsize K}}}      
\def\sfQ{\mbox{\sffamily{\scriptsize Q}}}      

\def\sbM{\mbox{{\bf \scriptsize M}}}
\def\sbN{\mbox{{\bf \scriptsize N}}}

\def\tc{\mbox{\tiny c}}

\def\te{\mbox{\tiny e}}

\def\tl{\mbox{\tiny l}}

\def\tm{\mbox{\tiny m}}

\def\bigupsigma{\mbox{\Large$\sigma$}}
\def\biguptau{\mbox{\Large$\uptau$}}
\def\bigtau{\mbox{\Large$\tau$}}

\def\K{Kucha\v{r} }

\def\pa{\partial}
\def\d{\textrm{d}}

\def\Circ{\mbox{\Large$\circ$}}               
\def\5Star{\mbox{\Large$\star$}}              
\def\cr{\mbox{\scriptsize{\bf $\mbox{ } \times \mbox{ }$}}}

\def\sumi2{\sum\mbox{}_{\mbox{}_{\mbox{\scriptsize $i$=1}}}^2}
\def\sumi3{\sum\mbox{}_{\mbox{}_{\mbox{\scriptsize $i$=1}}}^3}

\def\sumIN{\sum\mbox{}_{\mbox{}_{\mbox{\scriptsize $I$=1}}}^{N}}

\def\sumj3{\sum\mbox{}_{\mbox{}_{\mbox{\scriptsize $j$=1}}}^3}
\def\sumk3{\sum\mbox{}_{\mbox{}_{\mbox{\scriptsize $k$=1}}}^3}

\begin{document}

\begin{center}

{\large{\bf PROBLEM OF TIME: TEMPORAL RELATIONALISM COMPATIBILITY}} 

\vspace{0.08in}

{\large{\bf OF OTHER LOCAL CLASSICAL FACETS COMPLETED}} 

\vspace{.15in}

{\bf Edward Anderson} 

\vspace{.15in}

{\em DAMTP, Centre for Mathematical Sciences, Wilberforce Road, Cambridge CB3 OWA.  } 

\end{center}

\begin{abstract}

Temporal Relationalism (TR) is that there is no time for the universe as a whole at the primary level. 
Time emerges rather at a secondary level; one compelling idea for this is Mach's `time is to be abstracted from change'.
TR leads to, and better explains, the well-known Frozen Formalism Problem encountered in GR at the quantum level.
Indeed, abstraction from change is a type of emergent time resolution of this. 
Moreover, the Frozen Formalism Problem is but one of the many Problem of Time facets, which are furthermore notoriously interconnected.
The other local and already classically present facets are as follows.
2) The GR Thin Sandwich involves a subcase of Best Matching, which is one particular implementation of Configurational Relationalism.
3) The Constraint Closure Problem, 
4) the Problem of Observables or Beables, 
5) Spacetime Relationalism, 
6) the Spacetime Construction Problem, and 
7) the Foliation Dependence Problem as resolved in classical GR by Refoliation Invariance.
In this Article, I bring together the individual classical resolutions of these, and how these can be rendered compatible with TR.
Having covered that in detail for 2) to 6) elsewhere \cite{FEPI, TRiPoD, AM13}, 
%
%
the rest of the current Article is dedicated to the detailed form that 7) and its TR compatible modification takes. 
I.e. I consider TR implementing foliations, the TR versions of Refoliation Invariance 
and the associated TR version of hypersurface kinematics and hypersurface deformations.
These require a TR counterpart of the Arnowitt--Deser--Misner split.

\end{abstract}

\section{Introduction}\label{Introduction}

{\it Temporal Relationalism} (TR) is that there is no time for the universe as a whole at the primary level \cite{L}.
This can be implemented by a formulation of the following kind \cite{B94I, APoT3}.  

\mbox{ } 

\ni TR-i)  It is not to include any appended times -- such as Newtonian time -- or appended time-like variables such as the Arnowitt--Deser--Misner (ADM) \cite{ADM} lapse.  

\ni TR-ii) Time is not to be smuggled into the formulation in the guise of a label either.

\mbox{ }

\ni To formulate this, we need to know what {\sl does} remain available. 
This begins with the configurations $Q^{\sfA}$.
The collection of all possible values of the configuration of a physical system form the configuration space $\FrQ$ \cite{AConfig}.

\mbox{ } 

\ni In Sec 2 I lay out TR, with mathematical implementations and their consequences.
TR eventually leads to the well-known Frozen Formalism Problem facet 1) of the Problem of Time (PoT), alongside pointing to a resolution 
at the secondary level by Mach's \cite{M} `time is to be abstracted from change', i.e. a type of emergent time approach.
Sec 3 explains the other local PoT facets \cite{Kuchar92, I93, APoT2, APoT3} which are already present at the classical level. 
These are the 2) Thin Sandwich problem \cite{BSW, WheelerGRT, TSC1, TSC2}, which generalizes to the Best Matching problem \cite{BB82} and to Configurational Relationalism \cite{AGates}.
3) The Constraint Closure Problem, resolved for classical GR by the Dirac algebroid \cite{Dirac}.
4) The Problem of Observables or Beables \cite{ABeables}.
5) Spacetime Relationalism.  
6) Spacetime Construction \cite{RWR, AM13}. 
Finally, 7) the Foliation Dependence Problem, which is resolved for classical GR by the Dirac algebroid implying Refoliation Invariance \cite{Tei73}.  

\mbox{ }

\ni Sec 4 then explains how TR implementing (TRi) versions of all but one of the preceding have hitherto been constructed by use of TRiPoD (Principles of Dynamics) \cite{TRiPoD}.  
The current Article completes this by considering a TRi version of the theory of foliations \cite{IK85, I93}. 
This requires a more detailed account of GR as Geometrodynamics (Sec 5). 
This includes separating out 3-space, single hypersurface and multiple hypersurface or foliation concepts, 
alongside how the first two of these are dual interpretations of the same mathematical object. 
E.g. extrinsic curvature can be thought of as either a spatial tensor or a tensor on a hypersurface within spacetime, 
and Kucha\v{r}'s hypersurface derivative -- once viewed from the 3-space primary position -- becomes Barbour's Best Matching derivative.
I subsequently also replace ADM's split \cite{ADM} (Sec 6) by the TRi `A split' \cite{FEPI} (Sec 7).
Then some parts of the theory of foliations are outlined in Sec 8, and from a new starting point in Secs 9 and 10.
Sec 11 considers the TRi version of Refoliation Invariance, and Sec 12 of deformation first principles \cite{HKT}.
Sec 13 covers many-fingered time and its ray--wavefront dual, bubble time \cite{Bubble}.  
Sec 14 and the Conclusion cover consider how the Thin Sandwich contains a Spacetime Construction part as well as Configurational Relationalism parts, 
and can be extended to include recovery of Kucha\v{r}'s universal hypersurface kinematics \cite{Kuchar76I, Kuchar76II, Kuchar76III}. 
I also give there the TRi counterpart of this kinematics, which now admits a Machian interpretation.  

\vspace{10in}

\section{Temporal Relationalism (TR): implementations and consequences}\label{TR}

A first implementation of TR-ii) is for a label to be present but physically meaningless because it can be changed for any other (monotonically related) 
label without changing the physical content of the theory.   
E.g. at the level of the action, this is to be {\it Manifestly Reparametrization Invariant}.  
A finite-theory action of this kind is  

\ni\beq
\mS = 2 \int \d\lambda \sqrt{\mT \mW} \mbox{ } , \mbox{ } \mbox{ for } \mbox{ } 
\mT := ||\dot{\mbox{\boldmath{$Q$}}}||_{\sbM}\mbox{}^2/2 := \mM_{\sfA\sfB}\dot{Q}^{\sfA}\dot{Q}^{\sfB}/2 \mbox{ } .  
\label{S-Rel-Gen}
\eeq
$\mT$ is the the {\it kinetic energy}, $\bM$ the {\it configuration space metric} and $\mW = \mW(\mbox{\boldmath{$Q$}})$ is the {\it potential factor}.

\mbox{ } 

\ni I next introduce some model arenas which exemplify this; its consequences and further facets.  

\mbox{ } 

\ni Arena 1) Jacobi's action principle \cite{Lanczos-Arnol'd}, which can be considered to be for temporally-relational spatially-absolute Mechanics.  
Here $\FrQ = \mathbb{R}^{Nd}$ for $N$ particles in dimension $d$.
$\mT = ||\dot{\mbox{\boldmath$q$}}||_{\mbox{\scriptsize\boldmath $m$}}\mbox{}^2/2 = m_I\dot{q}^{iI}\dot{q}^{iI}/2$ so the configuration space metric $\mbox{\boldmath $m$}$ is just the 
`mass matrix' with components $m_{I}\delta_{IJ}\delta_{ij}$, and $\mW = E - \mV(\mbox{\boldmath $q$})$ for $\mV(\mbox{\boldmath{$q$}})$ the potential energy and $E$ is the total energy of the 
model universe.  

\ni Arena 2) Scaled 1-$d$ RPM with translations trivially removed \cite{FileR} is another case of Jacobi's action principle.
Here $\FrQ = \mathbb{R}^{nd}$ for $n := N - 1$ independent relative separation vectors
$\mT = ||\dot{\brho}||^2/2$ and $\mW = E - \mV(\brho)$ for $\brho$ the mass-weighted relative Jacobi coordinates \cite{Marchal-LR97, FileR} with components $\rho^{iA}$. 
This has the advantage over 1) of being a relational whole-universe model. 
See Sec 3.1 for a larger such model exhibiting further Background Independence (BI) aspects \cite{BB82}, and \cite{AMech} for an outline of a wider range of such models.  

\ni Arena 3) Full GR can indeed also be formulated in this manner.  
Here the the spatial 3-metrics $\mh_{ij}$ are the configuration variables, forming the configuration space $\mbox{Riem}(\bupSigma)$ on a fixed spatial topology $\bupSigma$.
This example involves an increasing number of departures of formalism from the familiar ADM action \cite{ADM}, though all of these formalisms remain equivalent.
For instance, the ADM lapse of GR is an appended time-like variable.
It can however be removed in the Baierlein--Sharp--Wheeler (BSW) action \cite{BSW, B94I}.
This has an indefinite kinetic term built using the inverse of the DeWitt supermetric \cite{DeWitt67} as kinetic metric, 
and potential factor $\mR - 2\, \slLambda$ for $\mR$ the spatial Ricci scalar and $\slLambda$ the cosmological constant.
See Sec 3.1 for the full form of this action.

\ni Arena 4) Misner's action principle \cite{Magic} is for the finite minisuperspace subcase of the previous example. 
In this case, $\mh_{ij}$ is the same at each spatial point.  
The above form for the potential carries over straightforwardly to the minisuperspace case. 
This model also retains GR's characteristic indefinite configuration space metric.  

\mbox{ } 

\ni The implementation of TR can be further upgraded as follows.
It is a further conceptual advance to formulate one's action and subsequent equations without use of any meaningless label at all.  
I.e. a {\it Manifestly Parametrization   Irrelevant} formulation in terms of {\it change}           $\d Q^{\sfA}$             rather than 
     a      Manifestly Reparametrization Invariant           one in terms of a label-time velocity  $\d Q^{\sfA}/\d \lambda$.

However, it is better still to formulate this directly: without even mentioning any meaningless label or parameter, 
by use of how the preceding implementation is dual to a Configuration Space Geometry formulation (\cite{Lanczos-Arnol'd} and c.f. the title of \cite{BSW}). 
Indeed, Jacobi's action principle is often conceived of geometrically \cite{Lanczos-Arnol'd} rather than in its dual aspect as a timeless formulation.
This formulation's action now involves not kinetic energy $T$ but a kinetic arc element $\d \ms$, or a physical Jacobi arc element $\d \mJ$: 

\ni\beq
\mS = \int \d \mJ = \sqrt{2} \int \d \ms \sqrt{\mW} \mbox{ } , \mbox{ } \mbox{ } 
\d \ms := ||\d \mbox{\boldmath$Q$}||_{\sbM} = \sqrt{\mM_{\sfA\sfB}(\mbox{\boldmath$Q$})\d Q^{\sfA}\d Q^{\sfB}} \mbox{ } . 
\label{S-Rel-Gen-2}
\eeq
See e.g. \cite{Lanczos-Arnol'd, BSW, FEPI, FileR, AM13} for how the product-type actions (\ref{S-Rel-Gen}, \ref{S-Rel-Gen-2}) 
are furthermore equivalent to the more familiar difference-type actions of Euler--Lagrange for Mechanics and of ADM for GR.  

\mbox{ }  

\ni The main way in which actions implementing ii) work is through their necessarily implying {\it primary constraints} \cite{Dirac}.  
I.e. relations between the momenta that are obtained without use of the equations of motion.
For (\ref{S-Rel-Gen}), this implication is via the following well-known argument of Dirac \cite{Dirac}. 
An action that is Manifestly Reparametrization Invariant is homogeneous of degree 1 in the velocities.  
Thus the $\mbox{dim}(\FrQ) = k$ conjugate momenta are (by the above definition) homogeneous of degree 0 in the velocities. 
Therefore they are functions of at most $k - 1$ ratios of the velocities. 
So there must be at least one relation between the momenta themselves (i.e. without any use made of the equations of motion).  
But this is the definition of a primary constraint.   
In this manner, TR acts as a constraint provider. 
See e.g. \cite{TRiPoD} for the counterpoint of this argument in the setting of (\ref{S-Rel-Gen-2}).

The constraint it provides has a purely quadratic form induced from \cite{B94I} that of the action (\ref{S-Rel-Gen-2})\footnote{`Purely' here  
means in particular that there is no accompanying linear dependence on the momenta.}  

\ni\beq
\scC\scH\scR\scO\scN\scO\scS :=  ||\mbox{\boldmath $P$}||_{\sbN}\mbox{}^2/2 - \mW( \mbox{\boldmath$Q$}) = 0   \mbox{ }  
\label{CHRONOS}
\eeq
for $\bN$ the inverse of the configuration space metric $\bM$. 
For instance, for Mechanics (\ref{CHRONOS}) is of the form 

\ni\beq
\scE := ||\mbox{\boldmath{$p$}}||_{\mbox{\scriptsize\boldmath{$n$}}}\mbox{}^2/2 + V(\mbox{\boldmath{$q$}}) = E \mbox{ } ,
\eeq 
for $\mbox{\boldmath{$n$}} = \mbox{\boldmath{$m$}}^{-1}$, with components $\delta_{IJ}\delta_{ij}/m_I$.  
This usually occurs in Physics under the name and guise of an {\it energy constraint}, though this is not the interpretation it is afforded in the Relational Approach (see below).
On the other hand, for GR it is the well-known and similarly quadratic Hamiltonian constraint $\scH$ -- now containing the DeWitt supermetric itself 
(see Sec 3 for its explicit form) that arises at this stage as a primary constraint.  

\mbox{ } 

\ni Reconciling timelessness for the universe as a whole at the primary level with time being apparent none the less in the parts of the universe that we observe,
proceeds via Mach's Time Principle \cite{M}: that `time is to be abstracted from change'. 
This gives a particular type of emergent time at the secondary level.
This Machian position is particularly aligned with the second and third formulations of TR ii), which are indeed in terms of change rather than velocity.
Classical emergent Machian times are of the general form

\ni\beq  
t^{\se\sm(\scc\sll)} = F[  Q^{\sfA}, \d  Q^{\sfA}] \mbox{ } . 
\label{t-Mach}
\eeq
A further issue is {\sl which} change. 
Some possibilities here are `any change' \cite{Rovelli} or `all change' \cite{B94I, Barbour}, 
though I have argued for the following third position intermediate between the preceding two \cite{ARel2, CapeTown, ABook}. 
A generalized local ephemeris time (GLET) is to be abstracted from a `sufficient totality of locally relevant change' (STLRC).
This builds in that some timestandards are more useful than others, while precluding problems with not knowing the contents of the universe well enough to include `all change'.  
This is a conception of time along the lines of the astronomers' ephemeris time \cite{Clemence}. 
(\cite{B94I, Barbour} also adopts this to some extent, though I take on board further aspects of it since the astronomical ephemeris clearly does not take `all change' into account. 
Also see \cite{KieferBook, Giu15} for further commentary.)

\mbox{ } 

\ni A specific implementation of GLET comes from rearranging (\ref{CHRONOS}).
This amounts to interpreting this not as an energy-type constraint but as an {\it equation of time} 
(thus termed $\scC\scH\scR\scO\scN\scO\scS$ after the Ancient Greeks' primordial God of Time). 
This rearrangement is aligned with 

\ni \beq 
\frac{\pa}{\pa t^{\se\sm(\scc\sll)}} = \frac{\sqrt{2W}}{\d s}    \frac{\pa}{\pa\lambda}
\eeq 
simplifying\footnote{Choosing 
time so that the equations of motion are simple was e.g. already argued for in \cite{MTW}.}
the system's momenta and equations of motion. 
Integrating this up, 

\ni \beq
t^{\se\sm(\scc\sll)} = \int \d s\left/\sqrt{2W}\right.  \mbox{ } \mbox{ {\it (emergent Jacobi time) }} . 
\label{t-em-J}
\eeq
\ni At the quantum level, (\ref{CHRONOS}) gives rise to   

\ni\beq
\widehat{\scC\scH\scR\scO\scN\scO\scS} \, \Psi = 0 \mbox{ } ,
\label{Q-CHRONOS}
\eeq
for $\Psi$ the wavefunction of the universe.
This includes the time independent Schr\"{o}dinger equation 

\ni\beq
\widehat{\scE}\Psi = E\Psi
\eeq
for Mechanics, and the well-known Wheeler--DeWitt equation  \cite{DeWitt67, Battelle} in the case of GR:

\ni\beq
\widehat{\scH}\Psi = 0 \mbox{ } . 
\eeq
Moreover, these equations occur in a situation in which one might expect time-dependent Schr\"{o}dinger equations 

\ni\beq
\widehat{H}\Psi = i\hbar\frac{\pa\Psi}{\pa t} \mbox{ } \left[ \mbox{ or } \mbox{ } i\hbar\frac{\delta\Psi}{\delta \mt}\right]
\eeq
for some notion of time $t$ [or $\mt$].  
Thus (\ref{Q-CHRONOS}) exhibits a {\it Frozen Formalism Problem} \cite{DeWitt67, Kuchar92, I93}.         
This is a well-known facet of the Problem of Time; on the other hand, 
the preceding steps which trace this back to the classical level and point to Machian resolutions are for now less well-known.

\mbox{ }

\ni (\ref{Q-CHRONOS})'s timelessness can be approached in an emergent semiclassical manner \cite{HH85, KieferBook}, 
which can furthermore also be interpreted in Machian terms \cite{ACos2, CapeTown}.  
As regards comparison of this with classical Machian emergent time, I first note that in the presence of a slow--fast and heavy--light ($h$--$l$) split 
-- modelling the small anisotropic and inhomogeneous corrections which enter cosmological calculations -- (\ref{t-Mach}) takes the form 

\ni\beq
t^{\se\sm(\scc\sll)} = F[h^{\sfA}, l^{\sfB}, \d h^{\sfA}, \d l^{\sfB}] \mbox{ } \underline{\sim} \mbox{ } F[h^{\sfA}, \d h^{\sfA}] \mbox{ } . 
\eeq
On the other hand, 

\beq  
t^{\se\sm(\sW\sK\sB)} = F[h^{\sfA}, \d h^{\sfA}, \upchi(h^{\sfA}, l^{\sfB}]] \mbox{ } \underline{\sim} \mbox{ } F[h^{\sfA}, \d h^{\sfA}] \mbox{ } 
\label{t-Mach-2}
\eeq
for $\Psi = \mbox{exp}(iS(h)/\hbar)\upchi(h, l)$ and $\upchi$ the wavefunction for the $l$-part of the system within which $h$ features as but a slowly-varying parameter.
Thus $t^{\se\sm(\scc\sll)}$ and $t^{\se\sm(\sW\sK\sB)}$ coincide to leading order, but not to greater accuracy than that. 
This difference is itself for a Machian reason \cite{ACos2}, namely that quantum change is not the same as classical change.


\section{The other Problem of Time facets}\label{PoT}

\subsection{Thin Sandwich, Best Matching and Configurational Relationalism}\label{CR}

{\it Configurational Relationalism} covers both 
\ni a) {\it Spatial Relationalism} \cite{BB82}: no absolute space properties.

\ni b) {\it Internal Relationalism} is the post-Machian addition of not ascribing any absolute properties to any additional internal space that is associated with the matter fields.

\mbox{ } 

\ni Configurational Relationalism involves the following postulates.

\mbox{ } 

\ni CR-i) One is to include no extraneous configurational structures either (spatial or internal-spatial metric geometry variables that are fixed-background rather than dynamical).

\ni CR-ii) Physics in general involves not only a $\FrQ$ but also a $\FrG$ of transformations acting upon $\FrQ$ that are taken to be physically redundant.
[To be clear, this is a {\it group action}: 
a map $\FrG \times \FrU \rightarrow \FrU$ for group $\FrG$ acting on some other mathematical space $\FrU$, presently a configuration space $\FrQ$.]

\mbox{ } 

\ni On some occasions, one can incorporate Configurational Relationalism through the availability of directly-invariant quantities.  
Such cases being few and far between, however, an indirect implementation along the lines of ii) is usually required.
A very general such is the `$\FrG${\it -act} $\FrG${\it -all}' method \cite{FileR, AGates}. 
Given an object $O$ that corresponds to the theory with configuration space $\FrQ$, one first applies a group action of $\FrG$ to this, denoted $\stackrel{\rightarrow}{\FrG}_gO$. 
This amounts to forming a $\FrG$-bundle version of $O$. 
Secondly, one applies some operation $\mbox{\Large S}_g$ that makes use of all of the $g^{\sfG} \in \FrG$ so as to cancel out the appearance of $g^{\sfG}$ in the group action. 
Examples of $\mbox{\Large S}_g$ then include summing, integrating, averaging, taking infs and sups, and extremizing, in each case over $\FrG$.  
{\it Group-averaging} in Group Theory and Representation Theory is perhaps the most obvious example.  
One can furthermore \cite{FileR, BI} insert `Maps' in between the $\FrG$-act and $\FrG$-all moves to make a general metric background invariantizing map

\ni\beq
\mbox{MBI} : \mbox{Maps} \circ O \mapsto O_{\sFG} = \mbox{\Large S}_{g \in \sFG} \circ \mbox{Maps } \circ \stackrel{\rightarrow}{\FrG} O \mbox{ } . 
\eeq
One of the more usually encountered realizations of the above indirect method is Barbour's {\it Best Matching} \cite{BB82}.\footnote{See e.g. 
\cite{Kendall, AGates, FileR} for further examples.}
%
This involves pairing $\FrQ$ with a $\FrG$ such that $\FrQ$ is a space of redundantly-modelled configurations.
Here $\FrG$ acts infinitesimally on $\FrQ$ as a {\it shuffling group}.
I.e. pairs of configurations are considered, with one kept fixed and the other shuffling around -- an active viewpoint -- until the two are brought into minimum incongruence.
$O$ is a Principles of Dynamics action and extremization over $\FrG$ for its $\FrG$-all operation.

\mbox{ } 

\ni It is helpful at this point to illustrate Best Matching by a simple example.  
Take for instance Euclidean RPM \cite{BB82, FileR}.  
This involves $\FrQ = \mathbb{R}^{Nd}$ of $N$ particles in dimension $d$, $\FrG = Eucl(d)$ and then the action 

\ni\be
\mS_{\sR\sP\sM} = 2\int \d\lambda \sqrt{\mW \, \mT_{\sR\sP\sM}}  \mbox{ } \mbox{ for } \mbox{ } 
\label{Jac} 
\mT_{\sR\sP\sM} = ||\Circ_{\underline{A}, \underline{B}}\mbox{\boldmath$q$}||_{\mbox{\scriptsize\boldmath$m$}}\mbox{}^2/2 
\mbox{ } \mbox{ and } \mbox{ } \Circ_{\underline{A}, \underline{B}} \, \mbox{\boldmath$q$} := \dot{\mbox{\boldmath$q$}} - \underline{A} - {\underline{B}} \cr \mbox{\boldmath$q$} \mbox{ }    
\ee
an example of Barbour's `{\it Best Matching derivative}'.\footnote{Here $\underline{A}$ is a translational auxiliary and $\underline{B}$ is a rotational one; 
$Eucl(d) = Tr(d) \rtimes Rot(d)$ for $\rtimes$ denoting semi-direct product.  
More generally in this Article I use underlines for spatial quantities, bolds for configuration space quantities and their conjugates, and overlining arrows for spacetime quantities.
I use slanty fonts for functions of one variable and straight fonts for functions of more than one variable.
I use round brackets for functions, square brackets for functionals and ( ; ] for mixed functional dependence.
I use small calligraphic font for constraints and small slanty font for observables or beables.}
%
Also, the potential involved is of the form $\mV(\mbox{\boldmath{$q$}}) = \mV(\underline{q}_I \cdot \underline{q}_J  \mbox{ alone})$. 
See Fig \ref{Facet-Intro-4}.d) for an example.

Then the momenta conjugate to the $\mbox{\boldmath{$q$}}$ are 

\ni \beq
\mbox{\boldmath{$p$}} = \sqrt{\mW/\mT_{\sR\sP\sM}} \, \{\dot{\mbox{\boldmath{$q$}}} - \underline{A} - \underline{B} \cr \mbox{\boldmath{$q$}}\} \mbox{ } .
\label{p-RPM}
\eeq
Then by virtue of the Manifest Reparametrization Invariance\footnote{While this is broken by the $- \underline{A} - \underline{B} \cr \mbox{\boldmath{$q$}}$ terms,
Dirac's argument extends over the breach, and the next Sec in any case lays out an unbroken replacement.} 
and the particular square-root form of the Lagrangian, the momenta obey a primary constraint quadratic in the momenta,  

\ni\be
\scE := ||\mbox{\boldmath $p$}||_{\mbox{\scriptsize\boldmath $n$}}\mbox{}^2/2 + V(\mbox{\boldmath $q$}) = E \mbox{ } . 
\label{calE}
\ee
Next, variation with respect to $\underline{A}$ and $\underline{B}$ give respectively the secondary constraints 

\ni\be
\underline{\scP} := \sumIN \underline{p}_{I} = 0 \mbox{ } \mbox{ } \mbox{ (zero total momentum constraint) } \label{ZM} \mbox{ } , \mbox{ } 
\ee

\ni\be
\underline{\scL} := \sumIN \underline{q}^{I} \cr \underline{p}_{I} = 0 \mbox{ } \mbox{ } \mbox{ (zero total angular momentum constraint) }
\label{ZAM} \mbox{ } .
\ee
These `shuffling constraints' are linear in the momenta. 
One then uses the inversion of the momentum--velocity relation (\ref{p-RPM}) to recast these in Lagrangian form, and eliminates $\underline{A}$ and $\underline{B}$ 
from these equations to obtain a reduced action \cite{FORD, FileR}.  
Attaining this entails passage to the quotient space $\FrQ/\FrG = \mathbb{R}^{Nd}/Eucl(d)$.

\mbox{ }

\ni Next, let us consider the geometrodynamical subcase of Best Matching, 
for which $\FrQ = \mbox{Riem}(\bupSigma)$ and $\FrG = Diff(\bupSigma)$: the corresponding diffeomorphisms. 
This case has close ties to Wheeler's so-called {\it Thin Sandwich}: Fig \ref{Facet-Intro-4}.b) and \cite{WheelerGRT, TSC1, TSC2, Kuchar92, I93}, 
thus indeed making contact with a second listed facet in the traditional PoT literature \cite{Kuchar92, I93}.  
This follows from the BSW action\footnote{$\mh_{ij}$ is the spatial 3-metric, with inverse $\mh^{ij}$ and determinant $\mh$. 
The overline denotes densitization, i.e. inclusion of a factor of $\sqrt{\mh}$. 
The corresponding 3-metric covariant derivative, Riemann tensor, Ricci tensor and Ricci scalar are $\mD_i$, ${\mR^i}_{jkl}$, $\mR_{ij}$ and $\mR$ respectively.  
Also $\cT$ is GR's kinetic term,                                               

\ni $\mM^{abcd} := \sqrt{\mh}\{\mh^{ac}\mh^{bd} - \mh^{ab}\mh^{cd}\}$ is the GR configuration space metric, whose inverse is the DeWitt supermetric \cite{DeWitt67} 

\ni $\mN^{abcd} :=           \{\mh^{ac}\mh^{bd} - \mh^{ab}\mh^{cd}/2\}/\sqrt{\mh}$.
$\dot{\mbox{ }} := \pa/\pa \lambda = \pa/\pa t$, since for GR label time $\lambda$ coincides with coordinate time $t$ on account of GR being an already-parametrized theory.  
$\pounds_{\underline{\upbeta}}$ is the Lie derivative with respect to the shift $\upbeta^i$.}

\ni\beq
\FS_{\sB\sS\sW} = \int \d t \int_{\Sigma}\d^3x \, \cL_{\sB\sS\sW} = \int\d\lambda\int_{\Sigma}\sqrt{\overline{\cT}} \sqrt{\overline{\mR} - 2\,\overline{\slLambda} \}} \mbox{ } , 
\overline{\cT} := \left| \left| \dot{\bh} - \pounds_{\underline{\upbeta}}\bh \right| \right|_{\sbM}^{\mbox{ }\mbox{ } 2} \mbox{ } . 
\label{S-BSW} 
\eeq 
One can arrive at this action from the more familiar ADM action

\ni \beq
\FS_{\sA\sD\sM} = \int \d t \int_{\Sigma}\d^3x \, \cL_{\sA\sD\sM} = \int \d t \int_{\Sigma}\d^3x\sqrt{\mh} \, \upalpha \{\mK_{ab}\mK^{ab} - \mK^2   +  \mR\} \mbox{ } .
\label{S-ADM}  
\eeq
Here, the {\it extrinsic curvature} $\mK_{ij}$ is related to the obvious Lagrangian variable $\dot{\mh}_{ij}$ by the computational formula

\ni\beq
\mK_{ij} = \frac{\dot{\mh}_{ij} -  \pounds_{\underline{\upbeta}}\mh_{ij}}{2 \, \upalpha} \mbox{ } ;   
\label{K-Comp-ADM}
\eeq
$\mK$ then denotes the trace of $\mK_{ij}$.  
The above-mentioned arrival is then by eliminating the ADM {\it lapse} $\upalpha$ from its own Lagrange multiplier equation: the Hamiltonian constraint

\ni\beq
\scH := ||\bp||_{\sbN}\mbox{ }^2 - \overline{\mR} + 2\, \overline{\slLambda} = 0 \mbox{ } .  
\eeq
recast in Lagrangian variables form $\cT/4\upalpha = \upalpha\{\mR - 2\, \slLambda\}$.
Trivial algebra then gives that $\upalpha = \frac{1}{2}\sqrt{\cT/\{\mR - 2\, \slLambda\}}$, and substituting this into (\ref{S-ADM}) produces (\ref{S-BSW}).  
Both of these actions can be set up in more detail upon Sec \ref{ADM}-\ref{A} more detailed considerations.
One can also postulate a (TRi version of) action (\ref{S-BSW}) from relational first principles \cite{B94I, RWR} as per the next Sec.
In this case, no lapse exists in the first place, with $\scH$ arising instead as a primary constraint.   
The Thin Sandwich is then the following extension of Wheeler's work \cite{BSW, WheelerGRT}. 

\ni Thin Sandwich 1) Consider the BSW action (\ref{S-BSW}) \cite{BSW}.  

\ni Thin Sandwich 2) Vary it with respect to $\upbeta^i$ \cite{BSW} to obtain the GR {\it momentum constraint}, best known in its Hamiltonian variables form  

\ni\beq
\scM_i := - 2 \, \mD_j p^j\mbox{}_i = 0 \mbox{ } .
\eeq
\ni Thin Sandwich 3) Now however solve the Lagrangian-variables form of $\scM_i$ (`thin sandwich equation') \cite{TSC1}  

\ni\be
\mD_{j}
\left\{ 
{\sqrt{\frac{2\, \slLambda - \mR(\ux; \bh]}{\{\mh^{ac}\mh^{bd} - \mh^{ab}\mh^{cd}\}
\{\pa{\mh}_{ab} - 2 \, \mD_{(a}\upbeta_{b)}\}
\{\pa{\mh}_{cd} - 2 \, \mD_{(c}\upbeta_{d)}\}  }}}
\{\mh^{jk} \delta^{l}_{i} - \delta^{j}_{i}\mh^{kl}\}\{\pa{\mh}_{kl} - 2 \, \mD_{(k}\upbeta_{l)}\}
\right\} 
= 0 \mbox{ } . 
\label{Thin-San}
\ee
with data ($\mh_{ij}, \dot{\mh}_{ij}$) for the shift $\upbeta^i(\ux) = \upbeta^i(\ux; \mh_{jk}, \dot{\mh}_{lm}]$.
In the sense of `PDE problem', this equation and data jointly constitute the Thin Sandwich Problem; Fig 1 explains the historical context of the `Thin Sandwich' name.  
%
{            \begin{figure}[ht]
\centering
\includegraphics[width=0.7\textwidth]{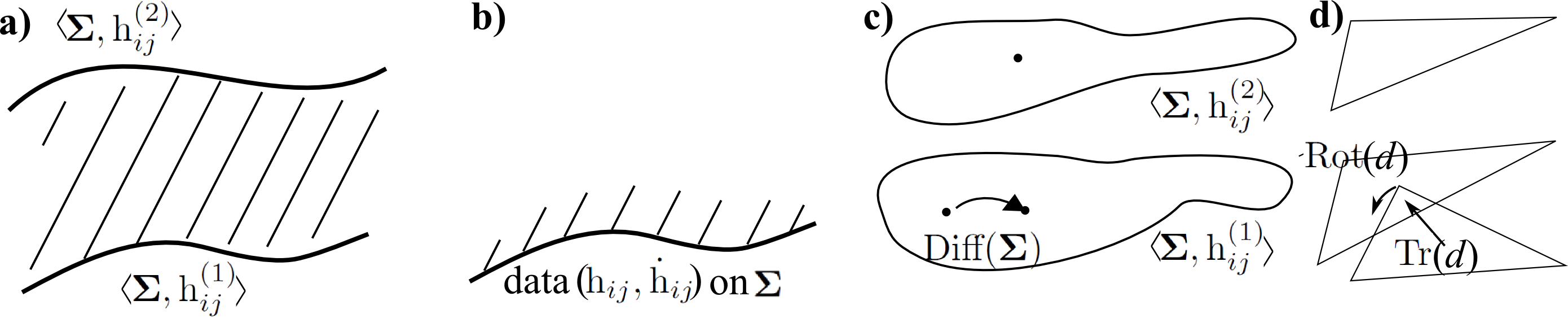}
\caption[Text der im Bilderverzeichnis auftaucht]{        \footnotesize{This is a historical account, 
whereas the text runs oppositely by starting with the most general case and specializing. 
a) \ni The `thin sandwich' name comes from Wheeler previously contemplating a `thick sandwich' \cite{WheelerGRT}: 
bounding bread-slice data $\mh_{ij}^{(1)}$ and $\mh_{ij}^{(2)}$ as knowns to solve for the spacetime `filling' in between. 
This was an attempted analogy with the quantum path integral of Feynman for transition amplitudes between states at two different times \cite{WheelerGRT}, 
but it failed to be mathematically well-posed.
b) Wheeler subsequently considered `thin sandwich' data to solve for a local coating of spacetime \cite{WheelerGRT}.
I.e. the `thin' limit of taking the data to be on the bounding `slices of bread', with data  $\mh_{ij}$ and $\dot{\mh}_{ij}$.
c) The Thin Sandwich can then be reinterpreted in terms of Best Matching $\mbox{Riem}(\bupSigma)$ with respect to $Diff(\bupSigma)$, 
which amounts to the shuffling as depicted \cite{B94I, RWR}.
I.e. keep one fixed  and shuffle the other to seek minimal incongruence between the two.  
See Sec \ref{A} for further details of this reinterpretation.
[I use $\langle \mbox{ } , \mbox{ } \rangle$ to indicate that a mathematical space is being equipped with further layers of mathematical structure.]
d) The analogous Best Matching for relational triangles with respect to the rotations $Rot(d)$ and translations $Tr(d)$ \cite{BB82, B94I, Lan}.
The triangle is in this case the minimal relationally nontrivial unit; \cite{AMech} tabulates such for a wider range of RPMs.
This subfigure is included to indicate Best Matching's applicability to a wider range of theories rather than just to geometrodynamics or the corresponding $Diff(\bupSigma)$.} }
\label{Facet-Intro-4} \end{figure}          }

\ni Thin Sandwich 4) Construct $\pounds_{\underline{\upbeta}}\mh_{ij}$: GR's $\vec{\mbox{Diff}}_{\underline{\upbeta}}\mh_{ij}$.   
Next construct Kucha\v{r}'s hypersurface derivative \cite{Bubble, Kuchar76II, Kuchar76III} 

\ni\beq
\Circ_{\vec{\upbeta}}\mh_{ij} := \dot{\mh}_{ij} - \pounds_{\vec{\upbeta}}\mh_{ij} \mbox{ } .
\label{Hyp-Der}
\eeq
Finally construct an emergent counterpart to $\upalpha$, $\mN := \mN(\ux; \mh_{ij}, \dot{\mh}_{kl}, \upbeta^m] =\sqrt{  \cT/4\{\mR - 2\, \slLambda\}  }$.

\ni Thin Sandwich 5) One can also substitute the $\upbeta^i$ obtained from 3) back in the original action to obtain a reduced action, 
thus completing Best Matching, and then regard this action as a new starting point.    

\ni Thin Sandwich 6) Thin Sandwich 3) and 4) also permit one to construct the extrinsic curvature $\mK_{ij} = \mK_{ij}(\ux; \mh_{lm}, \upbeta^n, \mN]$ \cite{BSW}. 
The formula for extrinsic curvature used here is (\ref{K-Comp-ADM}), except that BSW's {\sl emergent} $\mN$ has taken the place of ADM's {\sl presupposed} $\upalpha$.

\ni The Thin Sandwich Problem remains an open problem since the Thin Sandwich PDE's mathematics is hard \cite{TSC2}.

\subsection{Dirac Algorithm, Constraint Closure, and its Problems}\label{CC}

\ni Dirac's classification of constraints is into {\it first-class} constraints: those whose classical brackets with all the other constraints vanish weakly.  
{\it Second-class} constraints are then simply defined by exclusion to be those which are not first-class.  

\mbox{ } 

\ni Dirac begins to handle constraints by additively appending them with Lagrange multipliers to the incipient Hamiltonian for the system.
The {\it Dirac Algorithm} \cite{Dirac, HTbook} involves checking whether a given set of constraints implies any more constraints or any further types of entity.
The entities arising thus can be of five types.

\mbox{ } 

\ni Entity 0) {\it Inconsistencies} -- such as  $0 = 1$ arising as the Euler--Lagrange equations for $\mL = q$ \cite{Dirac}.  

\ni Entity 1) Mere {\it identities} -- equations that reduce to $0 \approx 0$, i.e. $0 = 0$ modulo constraints (Dirac's weak equality).

\ni Entity 2) Equations independent of the Lagrange-multiplier unknowns, which constitute extra secondary constraints.

\ni Entity 3) Demonstration that some of the constraints known so far are in fact second-class, by being second-class with respect to subsequently found constraints 
implied by the original constraints.

\ni Entity 4) Relations amongst some of the appending Lagrange multipliers functions themselves, which are a further `{\it specifier equation}' type of equation 
(i.e. specifying restrictions on the Lagrange multipliers).

\mbox{ }

\ni The Dirac Algorithm is to be applied recursively until one of the following three conditions holds.

\mbox{ }

\ni Termination 0) {\it Inconsistent theory} due to a case of 0) arising.

\ni Termination 1) {\it Trivial theory} due to the iterations of the Dirac Algorithm having left the system with no degrees of freedom. 

\ni Termination 2) {\it Completion}: the latest iteration of the Dirac Algorithm has produced no new nontrival consistent entities [2) to 4)], indicating that all of these have been found. 

\mbox{ } 
 
\ni The biggest Constraint Closure Problems are 0) alongside 1) when surviving degrees of freedom had been anticipated.  
However, there are other Constraint Closure Problems as well, such as being forced away from one's choice of group $\FrG$ by entity 2) arising, 
or finding that the shuffles one had anticipated to be gauge constraints were actually second-class by 3).    
These are $\FrG$-Constraint Closure Problems, some resolutions of which involve taking a smaller $\FrG$ or the larger $\FrG$ required by the integrabilities found.
See \cite{ABrackets} for more details.

\mbox{ } 

\ni For full GR, the algebraic structure formed by the constraints is\footnote{Here $[ \mbox{ } , \mbox{ } ]$ here denotes the geometrical commutator Lie bracket, 
whereas $X \overleftrightarrow{\pa}^i Y := \{ \pa^i Y \} X - Y \pa^i X$]: the 2-sided derivative familiar from QFT.}

\ni\be
\mbox{\bf \{} (    \scM_i    |    \upxi^i    ) \mbox{\bf ,} \, (    \scM_j    |    \upchi^j    ) \mbox{\bf \}} =  (    \scM_i    | \, [ \upxi, \upchi ]^i )  \mbox{ } ,
\label{Mom,Mom}
\ee

\ni\be
\mbox{\bf \{} (    \scH    |    \upzeta    ) \mbox{\bf ,} \, (    \scM_i  |    \upxi^i    ) \mbox{\bf \}} = (    \pounds_{\underline{\upxi}} \scH    |    \upzeta    )  \mbox{ } , 
\label{Ham,Mom}
\ee

\ni\be 
\mbox{\bf \{} (    \scH    |    \upzeta    ) \mbox{\bf ,} \,(    \scH    |    \upomega    )\mbox{\bf \}}  = (    \scM_i \mh^{ij}   |    \upzeta \, \overleftrightarrow{\pa}_j \upomega    )  \mbox{ } . 
\label{Ham,Ham}
\ee
Note foremost that this does close in the sense that there are no further constraints or other conditions arising in the right-hand-side expressions. 
Thus at the classical level for this Article's various toy models and full GR, the Constraint Closure Problem has the status of a {\sl solved} problem.

The first bracket means that $Diff(\bupSigma)$ on a given spatial hypersurface themselves close as an (infinite-dimensional) Lie algebra.
The second bracket means that $\scH$ is a good object -- a scalar density -- under $Diff(\bupSigma)$.  
Both of the above are kinematical rather than dynamical results.    
The third bracket, however, is more complicated in both form and meaning \cite{Tei73, Tei73b}.
The presence of $\mh^{ij}(\mh_{kl}(\ux))$ in its right hand side expression means the following.

\mbox{ } 

\ni i) The transformation itself to depend upon the object acted upon -- contrast with the familiar case of the rotations!  

\ni ii) The GR constraints to form a more general algebraic entity than a Lie algebra: a {\it Lie algebroid}.   
More specifically, (\ref{Mom,Mom}--\ref{Ham,Ham}) form the {\it Dirac algebroid} \cite{Dirac, BojoBook}.

\mbox{ } 

\ni By the Dirac algebroid, all of the above Constraint Closure issues are fine in the case of classical GR.

\subsection{Expression in terms of Beables or Observables, and its Problems}\label{PoB}

Having found constraints and introduced a classical brackets structure, one can then ask which objects have zero classical brackets with one's constraints: 

\ni\beq
\mbox{\bf |[} \scC_{\sfA} \mbox{\bf ,} \, \iB_{\sfB} \mbox{\bf ]|} \mbox{ } `=' 0 \mbox{ } .
\eeq
These objects -- usually termed {\it observables} -- 
are more physically useful than just any functions (or functionals) of $Q^{\sfA}$ and $P_{\sfA}$, due to their containing physical information only.  
Note that this notion allows for multiple possible kinds of brackets ($\mbox{\bf |[} \mbox{ } \mbox{\bf ,} \mbox{ } \mbox{\bf ]|}$ being a generic such), 
multiple possible notions of equality (most commonly Dirac's $\approx$) and multiple possible sets of constraints.
However, the Jacobi identity applied to two constraints and one observable requires that the input notion of constraints is a closed algebraic structure \cite{ABeables}. 
Applied instead to one constraint and two observables, it establishes that observables themselves form a closed algebraic structure.  
In this sense, observables form an algebraic structure that is associated with that which is formed by the constraints themselves.

I also use an extension from the notion of observables, which eventually carry nontrivial connotations of `are observed', to {\it beables}, which just `are'.\footnote{The latter
-- a term originally coined by Bell \cite{Bell} -- are somewhat more general. 
This generalization does not concern a change of definition but rather a more inclusive context in which the entities are interpreted.
I.e. beables accommodate a wider range of viable realist approaches, by which I mean specifically consistent histories \cite{Hartle}, 
approaches based on decoherence \cite{GiuBook} and Isham and Doering's topos-based contextual realism \cite{ID}. 
It so happens to also accommodate Bohmian conceptualization for those who have a taste for that.
But the previous three cases are the ones due to which I specifically use the beables concept, and those can be considered entirely freely from any Bohmian premises.
Nor is any of what I specifically accommodate here related to un-viable realist approaches such as overruled types of hidden variables theory.
Beables retain a smaller amount of nontrivial meaning in the classical {\sl whole-universe} setting, and then whole-universe Quantum Cosmology combines these two elements. 
Also note here that this causes of consistent histories and decoherence to differ in conceptual character between subsystem/laboratory setting and whole-universe Quantum Cosmology setting.
So one should not assume in reading these terms that one is familiar with their manifestation in Quantum Cosmology, unless one has studied those explicitly \cite{GiuBook}.}

In particular then, {\it Dirac observables or beables} \cite{DiracObs} are quantities that (for now classical) brackets-commute with all of a given theory's first-class constraints, 

\ni\beq
\mbox{\bf \{}    \scC_{\sfC}    \mbox{\bf ,} \, \iD_{\sfD}   \mbox{\bf \}} \approx 0 \mbox{ } .
\eeq 
On the other hand, \K introduced \cite{Kuchar93} another type of observables, which were indeed subsequently termed \K observables 
(and generalize in the above wider context to what I term {\it \K beables}). 
These are quantities which form zero brackets with all of a given theory's first-class linear constraints, 

\ni\beq
\mbox{\bf \{}    \scF\scL\scI\scN_{\sfI}    \mbox{\bf ,}  \, \iK_{\sfK}    \mbox{\bf \}} \approx 0 \mbox{ } .
\eeq
Specific examples of classical \K observables or beables are that these are trivially any quantities of the theory for minisuperspace and spatially-absolute mechanics. 
Less trivially, for RPMs more generally these include shapes and scales \cite{ABeables2}.
\K observables or beables coincide with $Diff(\bupSigma)$ {\it gauge-invariant quantities} in GR, for which the condition for them is 

\ni\beq
\mbox{\bf \{}    (\scM_{i}|\upxi^i)    \mbox{\bf ,}  \, (\iK_{\sfK}|\uptheta^{\sfK})    \mbox{\bf \}} \approx 0 \mbox{ } . 
\eeq
e.g. the configurational subcases of which is solved formally by the 3-geometries themselves.  
The extra condition in GR for the above to be additionally Dirac is then that

\ni\beq
\mbox{\bf \{}    (\scH|\upzeta)    \mbox{\bf ,}  \, (\iK_{\sfK}|\upphi^{\sfK})    \mbox{\bf \}} \approx 0 \mbox{ } .  
\eeq
\ni Finally, the {\it Problem of Beables} or {\it Observables} is that it is hard to construct a set of beables, in particular for Gravitational Theory.  
More specifically, Dirac observables or beables are harder to find than \K ones, and each's quantum counterparts are even harder to find than classical ones.   
See e.g. \cite{H03-AHall} for some model arena work on constructing Dirac observables or beables.

\subsection{Spacetime Relationalism and its classical lack of Problems}\label{SRel}

\ni STR-i) There are to be no background spacetime structures, in particular no indefinite-signature background spacetime metrics. 
Fixed background spacetime metrics are also more well-known than fixed background space metrics. 

\ni STR-ii) Now as well as considering a spacetime manifold $\FrM$, consider also a $\FrG_{\sS}$ of transformations acting upon $\FrM$ that are taken to be physically redundant.

\mbox{ } 

\ni For GR,  $\FrG_{\sS} = Diff(\FrM)$.
\ni Now $\FrM$ can be equipped with matter fields in addition to the metric. 

\ni Additionally it has its own closure: $Diff(\FrM)$ forms a Lie algebra

\ni \be
\mbox{\bf |[} 
(    \mD_{\mu}    |    \upeta^{\mu})    
\mbox{\bf ,} \, 
(    \mD_{\nu}| \upiota^{\nu}    ) 
\mbox{\bf ]|} = 
(    \mD_{\gamma}    | \, [\upeta, \upiota]^{\gamma}    ) \mbox{ } \label{Lie-2} . 
\ee

for $\mD_{\mu}$ here the generators of $Diff(\FrM)$.  
It additionally has its own notion of observables: quantities $\mS_{\sfQ}$ such that

\ni\be
\mbox{\bf |[} (    \mD_{\mu}    |    \upeta^{\mu})  \mbox{\bf ,} \, (    \mS_{\sfQ}| \uppsi^{\sfQ}    ) \mbox{\bf ]|} \mbox{ }  `=' \mbox{ } 0 \mbox{ } \label{Sp-Obs} 
\ee
Thus `where observables fit' in programs addressing the Problem of Time is partly resolved by there being multiple notions of observables which fit in multiple places!

\subsection{Foliation Independence classically attained Refoliation Invariance}\label{FI-RI}

{\it Foliation Dependence} is a type of privileged coordinate dependence. 
This runs against the basic principles of what GR contributes to Physics.


\ni {\it Foliation Independence}       is then an aspect of BI, and the 
    {\it Foliation Dependence Problem} is the corresponding Problem of Time facet if the preceding fails to be met.  
It is obviously a time problem since each foliation by spacelike hypersurfaces is orthogonal to a GR timefunction.
I.e. each slice corresponds to an instant of time for a cloud of observers distributed over the slice.
Each foliation corresponds to the cloud of observers moving in a particular way (Fig \ref{Observer-Congruences} develops this further).   
GR necessitates this diversity of observers, and the corresponding foliations experienced as sequences of instances by these 
explain the passage, upon splitting spacetime's $Diff(\FrM)$, to the much larger Dirac algebroid.


\ni The {\it Refoliation Invariance} question is then posed in Fig \ref{Refol-4-SCPs}.a) and affirmed in Fig \ref{Refol-4-SCPs}.b) in the case of classical GR. 
%
{            \begin{figure}[ht]
\centering
\includegraphics[width=0.5\textwidth]{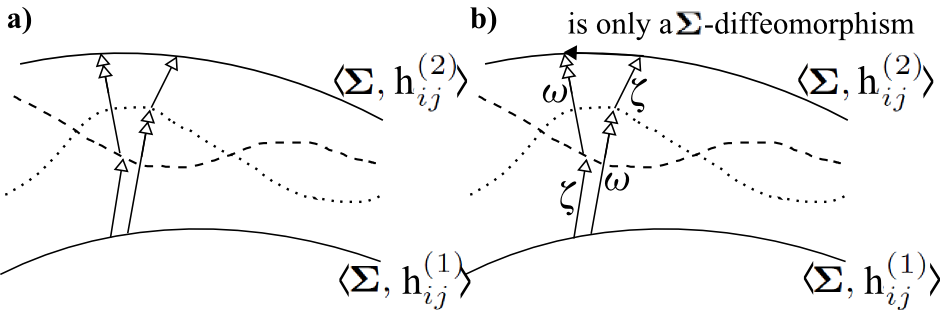}
\caption[Text der im Bilderverzeichnis auftaucht]{\footnotesize{a) The Foliation Dependence Problem is whether evolving between 
initial $\langle \bupSigma, \bh^{(1)} \rangle$ and final $\langle \bupSigma, \bh^{(2)} \rangle$ spatial hypersurfaces 
via each of the dashed and the dotted spatial hypersurfaces give the same physical answers. 
b) Teitelboim and Kucha\v{r}'s \cite{Tei73, Bubble} (further exposited in \cite{Kuchar73, HKT, Teitelboim} classical `Refoliation Invariance' resolution affirms this: 
the Poisson bracket (\ref{Ham,Ham}) within the Dirac algebroid leads to the two evolutions differing only by a diffeomorphism within the final hypersurface.
This constitutes the `{\it Refoliation Invariance Theorem of GR}'.} }
\label{Refol-4-SCPs} \end{figure}          }

\subsection{Spacetime Construction from space classically attained}\label{SCP}

Next consider assuming less structure than spacetime, with the aim of recovering spacetime, at least in a suitable limit. 
This can be hard, particularly in approaches assuming a lesser amount of structure.
This aspect was originally known as `Spacetime Reconstruction', though I take that to be too steeped in assuming spacetime primality and thus rename it `Spacetime Construction'.
Wheeler motivated such ventures as follows.

\mbox{ }

\ni 1) At the quantum level, fluctuations of the dynamical entities are inevitable. 
In the present case, as Wheeler pointed out \cite{Battelle} these are fluctuations of 3-geometry; these are then too numerous to be embedded within a single spacetime.  
The beautiful geometrical way that classical GR manages to be refoliation invariant breaks down at the quantum level.

\mbox{ } 

\ni 2) Heisenberg's uncertainty principle is furtherly relevant \cite{Battelle, W79}.
Precisely-known position $\underline{q}$ and momentum $\underline{p}$ for a particle are a classical concept corresponding to a sharp worldline.
This view of the world is entirely accepted to break down in quantum physics due to Heisenberg's Uncertainly Principle.
In QM, worldlines are replaced by the more diffuse notion of wavepackets. 
However, in GR, what the Heisenberg uncertainty principle now applies to are the quantum operator counterparts of $\mh_{ij}$ and $\mp^{ij}$. 
But by formula (\ref{Gdyn-momenta}) this means that $\mh_{ij}$ and $\mK_{ij}$ are not precisely known.   
Thus the idea of embeddability of a 3-space with metric $\mh_{ij}$ within a spacetime is itself quantum-mechanically compromised.

\mbox{ } 

\ni 3) Wheeler \cite{Battelle} asked the following question, which readily translates to asking for first-principles reasons for the form of the 
crucially important GR Hamiltonian constraint, $\scH$.
\beq
\stackrel{\mbox{\it \normalsize ``If one did not know the Einstein--Hamilton--Jacobi equation, how might one hope to derive it straight off from}}  
         {\mbox{\it \normalsize       plausible first principles without ever going through the formulation of the Einstein field equations themselves?"}}		 
\label{Wheeler-Q}
\eeq
Here, one is no longer thinking just of GR's specific geometrodynamics, but rather of a multiplicity of geometrodynamical theories.  
One is then to look for a {\sl selection principle} that picks out the GR case of geometrodynamics.

\mbox{ } 

\ni Let us for now concentrate on Spacetime Construction from assuming space.
This goes beyond Hojman, \K and Teitelboim's (HKT) \cite{HKT} first answer to Wheeler's question 
(which we also have occasion to visit in Sec \ref{HKT}) which assumes embeddability into spacetime.  
The second answer \cite{RWR, AM13}, however, does not assume spacetime, being based rather on 3-spaces $\bigupsigma$ in place of hypersurfaces $\bupSigma$; 
it proceeds from Temporal and Configurational Relationalism first principles.
Then the consistency of the ensuing constraints' algebraic structure along the lines of the Dirac Algorithm returns.
This arises from a more general\footnote{Here $\mN_{abcd}^{x,y} := y\{  \mh_{ac}\mh_{bd} -  x \,\mh_{ab}\mh_{cd}/2 \}/\sqrt{\mh}$
is the inverse of $\mM^{abcd}_{w,y} := \sqrt{\mh}\{\mh^{ac}\mh^{bd} - w \, \mh^{ab}\mh^{cd}\}/y$ for $x := 2w/\{3 w - 1\}$. \label{MN-Ansatze}}  

\ni\beq
\scH_{\st\sr\si\sa\sll} = \scH_{x,y,a,b} := ||\bp^{ab}||_{\sbN^{x, y}}\mbox{}^2 - \sqrt{\mh}\{a \, \mR + b\} = 0 \mbox{ } , 
\eeq
in the sense that 

\ni\beq
\mbox{\bf \{} (    \scH_{x,y,a,b}    |    \upzeta    ) \mbox{\bf , } \, (    \scH_{x,y,a,b}    |    \upomega    ) \mbox{\bf \}}
\eeq
gives, aside from terms weakly vanishing due to already-known constraints, the 4-factor obstruction term \cite{RWR, Van, AM13}

\ni\be
2 \times a \times y \times \{ 1 - x \} \times (    \mD_i \mp \, | \, \upzeta \, \overleftrightarrow{\pa}^i \upomega    ) \mbox{ } .  
\label{4-Factors}
\ee 
Then setting the second factor to strongly be zero corresponds to recovery of GR's DeWitt supermetric, so the Dirac Algorithm succeeds in retrieving GR. 
In considering minimally-coupled matter as well, a further obstruction term arises with a factor corresponding to local Lorentzian relativity ($c$ finite, null cones).  
Moreover, due to the other factors present in the obstruction terms, GR is not recovered alone. 
$y = 0$ gives a geometrostatics with (a generalization of \cite{Lan2, AM13}) Galilean Relativity ($c = \infty$, null cones squashed into planes). 
$a = 0$ gives Strong Gravity \cite{I76}, with matching Carrollian Relativity \cite{LL} ($c = 0$, null cones squeezed into lines). 
Thus choice of relativity here becomes captured by mathematically satisfying Constraint Closure by strong equality in the Dirac Algorithm.  
[Contrast the form of Einstein's historical dichotomy of Galilean versus Lorentzian universal Relativity holding locally.]  
The final factor can be satisfied weakly by $\mD_i\{\mp/\sqrt{\mh}\} = 0$.
In particular, maximal $\mp = 0$ and constant mean curvature $\mp/\sqrt{\mh} = const$ conditions solve this, 
leading to various conformogeometrodynamical formulations of GR \cite{ABFKO, BO10} and alternative conformogeometrodynamical theories \cite{ABFO, GGK11}.   
That is not an option that the current Article pursues any further, concentrating rather on completing the most orthodox emergent GR as geometrodynamics case above.  

\mbox{ }

\ni Other aspects of Spacetime Construction are then effectuated by Thin Sandwich 6)'s construction of extrinsic curvature 
and by the above recovery of the Dirac algebroid being reinterpretable as embeddability into GR's notion of spacetime \cite{Tei73}. 

\mbox{ } 

\ni The above completes the {\it local} classical description of the aspects of Background Independence with ensuing Problem of Time facets arising from those which are not met.
See e.g. \cite{Kuchar92, I93, AGlob} for some of the large number of further Global Problems of Time, both as regards time properties and as regards facets coming with global issues.
See e.g. \cite{Kuchar92, I93, APoT2, APoT3, ABook} (or \cite{APoT, CapeTown} for summaries) for quantum aspects of Background Independence and ensuing Problem of Time facets.
The only main facet which is not realized until the quantum level are the so-called Multiple Choice Problems \cite{I93, Gotay, FileR}.

\section{TR implementing (TRi) versions of the other classical facets}\label{TRiPoT}

{            \begin{figure}[ht]
\centering
\includegraphics[width=0.9\textwidth]{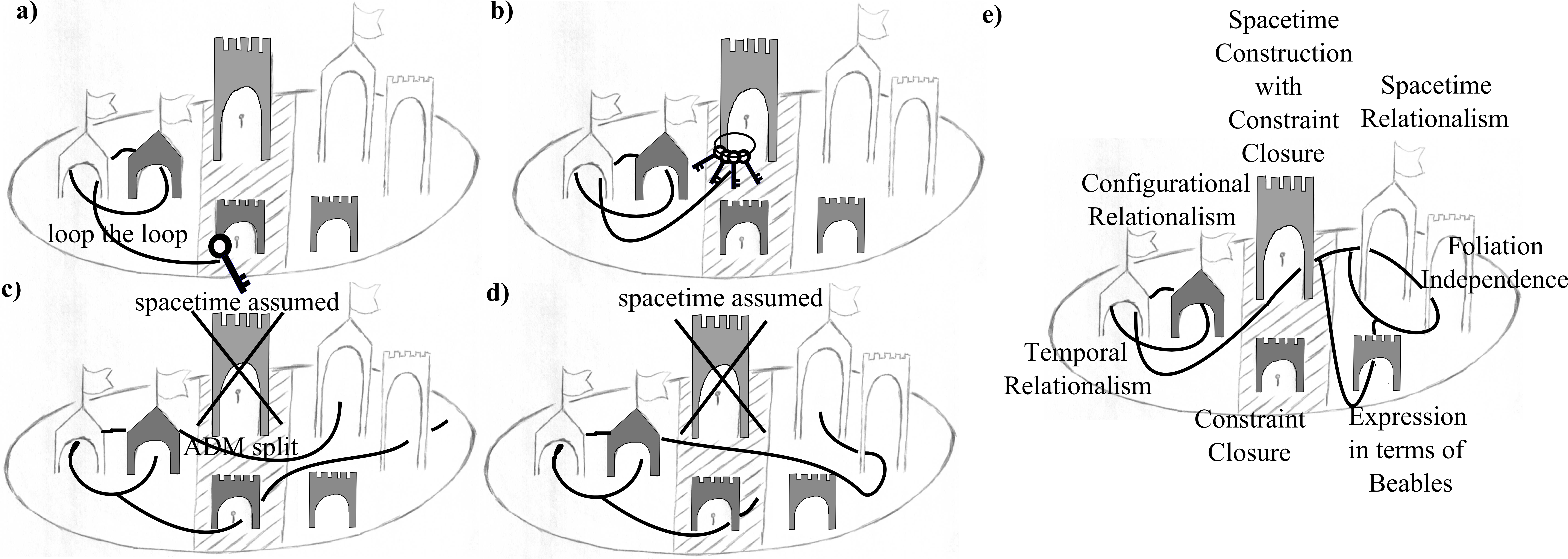}
\caption[Text der im Bilderverzeichnis auftaucht]{\footnotesize{Various orderings of the classical gates \cite{AGates} which depict the local Problem of Time facets as per e).

\ni a) depicts the Configurational and Temporal Relationalism `loop the loop' of Secs \ref{TR} and \ref{CR}, 
by which a geometrical best-matched action $\FS$ is produced on a configuration space $\FrQ$ for which the group $\FrG$ of transformations is held to be physically irrelevant.
The candidate triple $\FrQ$, $\FrG$, $\FS$ produces constraints $\scC\scH\scR\scO\scN\scO\scS$ and $\scS\scH\scU\scF\scF\scL\scE_{\sfG}$. 
These are a candidate key with which to attempt to open the Constraint Closure gate, in accord with the TRiPoD version of Dirac's algorithm.   
In the case of GR, $\scH$ and $\scM_i$ indeed close, by the Dirac algebroid (in its TRiPoD-smeared version in the Relational Approach).

\ni b) repeats the above with a family of theories to produce a bunch of keys.
Using ansatz (\ref{trial}), one of these recovers the GR form of $\scH$, attains Spacetime Construction, and recovers locally SR physics in addition to obeying Constraint Closure.  

\ni [In attempting each of a) and b), if one's $\FrQ$, $\FrG$, $\FS$ triple(s) fails to produce any keys, one is to go back to the loop and try again.]

\ni c) outlines the order in which the gates are approached in Secs \ref{Gdyn} and \ref{ADM}.
Here spacetime is assumed ab initio. 
The ADM split of the metric then leads to $\scH$ and $\scM_i$ arising together from the corresponding split of the Einstein--Hilbert action and its subsequent variation. 
[Moreover, one could treat $\scM_i$ second here, since it is an integrability of $\scH$, ordering rather than splitting this path.]
Constraint Closure is then attained as per the usual form of the Dirac algebroid.
The A split version of Sec \ref{A} also parallels this ordering.

\ni d) outlines the order in which the gates are approached in the foliations-first parts of Secs \ref{Fol-2}-\ref{Hyp-Kin}.
Here spacetime assumed ab initio and then foliations are established.
This split of spacetime produces the Dirac algebroid in the form of the deformation algebroid.
One then deduces that the generators of this must take the forms of $\scM_i$ and $\scH$ under various assumptions (Sec \ref{HKT}).

\ni e) Is finally the current Article's main route from relational first principles to emergent spacetime constructed and then its subsequent foliation.} }  
\label{Gates-2} \end{figure}          }

\ni These are ordered as in the branching path of Fig \ref{Gates-2}.a) which also contrasts this with various other gate orderings.  
This ordering is a major issue since the Problem of Time facets have been likened to the gates of an enchanted castle \cite{Kuchar93}. 
I.e. if one goes through one of them and then through another, one often finds oneself once again outside of the first one. 
The Figure considers only the {\sl classical} version of this `gates problem'.

\subsection{Temporal and Configurational Relationalism combined}\label{TRi-CR}

Placing Configuration Relationalism implemented by Lagrange multipliers into a Manifestly Reparametrization Invariant implementation of TR breaks the latter implementation.  
One can get around this by implementing Configuration Relationalism using cyclic velocities $\dot{g}$ instead; 
free end notion of space variation ensures this is equivalent to the preceding. 
Moreover, the geometrical dual to Manifest Parametrization Irrelevance is a superior implementation of TR. 
In this case then Configuration Relationalism is implemented using cyclic differentials $\d g$ instead.  
Actions are then of the form 

\ni\beq
\mS =  \int \d \mJ = \sqrt{2} \int \sqrt{\mW}\d \ms \mbox{ } , \mbox{ } \mbox{ }
\d \ms := ||\d_{g}\mbox{\boldmath$Q$}||_{\mbox{\scriptsize\boldmath$M$}} 
\mbox{ } \mbox{ for } \mbox{ } \d_{g}\mbox{\boldmath$q$} := \d{\mbox{\boldmath$q$}} - \stackrel{\rightarrow}{\FrG}_g \mbox{\boldmath$Q$} \mbox{ } 
\eeq
For instance, for Euclidean RPM, 

\ni\beq
\mS_{\sR\sP\sM} = \sqrt{2} \int \sqrt{\mW}\d \ms_{\sR\sP\sM} \mbox{ } , \mbox{ } \mbox{ }
\d \ms_{\sR\sP\sM} := ||\d_{\underline{A}, \underline{B}}\mbox{\boldmath$q$}||_{\mbox{\scriptsize\boldmath$m$}} 
\mbox{ } \mbox{ for } \mbox{ } \d_{\underline{A}, \underline{B}}\mbox{\boldmath$q$} := \d{\mbox{\boldmath$q$}} - \d\underline{A} - {\d\underline{B}} \cr \mbox{\boldmath$q$} \mbox{ } .    
\label{Jac-2} 
\eeq
A new TRiPoD definition of momentum is now required \cite{FileR, TRiPoD}: 

\ni\beq
P_{\sfA} := \frac{\pa \, \d \mJ}{\pa \, \d Q^{\sfA}}
\eeq
for the general finite-theory case.
Then the Best Matching Problem becomes a Jacobi--Mach level elimination, though mathematically it has the same form as previously.
Also note how Best Matching needs to be resolved before $t^{\se\sm(\scc\sll)}$ becomes explicit:  

\ni\beq
t^{\se\sm(\scc\sll)} = \mbox{\Large E}_{\sg \in \FrG} \int \d \ms_{\sg}/\sqrt{2\mW}  \mbox{ } 
\label{t-em-J-G}
\eeq
this extremization (denoted by $\mbox{\Large E}$: a subcase of $\FrG$-all's $\mbox{\Large S}$) acting so as to wipe out the $\FrG$-dependence.
Furthermore, this extremization is of a second functional -- the action itself -- rather than of the obvious first functional in (\ref{t-em-J-G}); 
in general extremizing over the first functional itself is inconsistent \cite{FileR}.  

\mbox{ } 

\ni Next let us pass over to field theory (GR in particular). 
The fully relational action for GR is  

\ni\be
\FS_{\sr\se\sll} = \int \int_{\Sigma}\d^3 x \cJ  =  \int \int_{\Sigma}\d^3 x  \sqrt{\overline{\mR} - 2\, \overline{\slLambda}}\,\pa \cs  \mbox{ } , \mbox{ } \mbox{ } 
\pa \cs := \left|\left|\pa_{\underline{\sF}}\bh\right|\right|_{\sbM} \mbox{ } . 
\label{S-Rel}
\ee
Here $\pa\cJ$ stands for the Jacobi arc element density and $\pa\cs$ for kinetic arc element half-density.
$\pa_{\underline{\sF}}\mh_{ij} := \pa \mh_{ij} - \pounds_{\pa \underline{\sF}} \mh_{ij}$ is the Best Matching corrected derivative for GR, 
and $\pa \mF^i$ is the {\it differential} of the {\it frame variable}, $\mF^i$
\footnote{Another route to this action is from the A action (\ref{S-A}) by the cyclic differential analogue of Routhian reduction \cite{TRiPoD}.
This is the TRi counterpart of Sec 2's ADM to BSW move by Lagrange multiplier elimination.}
%
The conjugate momenta are then 

\ni\beq
\mp^{ij} :=  \frac{  \delta \, \pa \cJ  }{  \delta \, \pa \mh_{ij}  } = \mM^{ijkl} 2\sqrt{ \overline{\mR} - 2\, \overline{\slLambda} } \frac{\pa_{\underline{\sF}}\mh_{kl}}{\pa \cs} \mbox{ } . 
\label{tumvel}
\eeq
Then $\scH$ and $\scM_i$ follow more or less as before. 
$\scH$ arises from the following uplift of Dirac's argument to Jacobi--Mach form as follows \cite{TRiPoD}. 
A geometrical action is dual to a Manifestly Parametrization Irrelevant action. 
Thus it is homogeneous of degree 1 in its changes. 
So each of its total of $\mbox{dim}(\FrQ) = k$ momenta are homogeneous of degree 0 in the changes.
Thus these are functions of $k$ -- 1 independent ratios of changes. 
So there must be at least one relation between the momenta themselves without any use made of the equations of motion.  
But by definition, this is a primary constraint, so we are done.
On the other hand, $\scM_i$ arises as a secondary constraint from free end spatial hypersurface variation with respect to the auxiliary $Diff(\bupSigma)$-variables $\mF^i$

\mbox{ } 

\ni I next reiterate the Thin Sandwich procedure in the TRiPoD formulation's manifestly Machian terms.

\mbox{ }

\ni Machian Thin Sandwich 1) Consider the relational GR action (\ref{S-Rel}) \cite{RWR, AM13}.  

\ni Machian Thin Sandwich 2) Vary it with respect to $\mF^i$ to obtain $\scM_i$ \cite{ABFO}.

\ni Machian Thin Sandwich 3) Solve the Jacobi variables form of $\scM_i$ \cite{AHH}
\ni\beq
\mD_{j}\left\{ {\sqrt{\frac{2\slLambda - \mR(\ux; \bh]}{\{\mh^{ac}\mh^{bd} - \mh^{ab}\mh^{cd}\}\{\pa{\mh}_{ab} - 2\mD_{(a}\pa\mF_{b)}\}\{\pa{\mh}_{cd} - 2\mD_{(c}\pa\mF_{d)}\}  }}}
                                                      \{\mh^{jk} \delta^{l}_{i} - \delta^{j}_{i}\mh^{kl}\}\{\pa{\mh}_{kl} - 2\mD_{(k}\pa{\mF}_{l)}\}\right\} = 0 \mbox{ } . 
\label{TR-Thin-San}
\eeq
with Machian data ($\mh_{ij}, \pa {\mh}_{ij}$) for the differential of the frame auxiliary $\pa\mF^i$.   
Note that altering (\ref{Thin-San}) to (\ref{TR-Thin-San}) makes no difference to the {\sl mathematical form} of the PDE problem.  
							 						 
\ni Machian Thin Sandwich 4a) Construct $\pounds_{\pa\underline{\sF}} \mh_{ij}$, and then the Best Matching derivative   

\ni\beq
\pa_{\underline{\sF}}\mh_{ij} = \pa \mh_{ij} - \pounds_{\pa\underline{\sF}}\mh_{ij} \mbox{ } . 
\label{BM-Der}
\eeq
This is a distinct conceptualization of the same mathematical object as the hypersurface derivative (see Sec \ref{A} for a further account of this).
Then construct the {\it emergent differential of the instant} 

\ni\beq
\pa\mI = ||\pa_{\underline{\sF}} \bh||_{\sbM}  \big/  2\sqrt{ \overline{\mR} - 2 \, \overline{\slLambda}  }  \mbox{ } .
\label{gue}
\eeq
\ni Machian Thin Sandwich 4b) An emergent time readily follows simply from integrating up 4a) \cite{Christodoulou, B94I, RWR, AM13}. 
Note that 4b) goes beyond BSW's own construction.
It is GR's analogue of emergent Jacobi time as highlighted by Barbour \cite{B94I}.
It is an `all change', or, in practise `STLRC' implementation of Mach's Time Principle: 
\ni\beq
\mt^{\se\sm(\scc\sll)}(\ux) = \mbox{\large E}_{\underline{\sF} \mbox{ }  \in  \mbox{ }  \mbox{\scriptsize Diff}(\Sigma)        }
\int \left. ||\pa_{\suF}\bh||_{\sbM} \right/   \sqrt{     \overline{\mR} - 2\overline{\slLambda}  }   \mbox{ } .
\label{GRemt2}
\eeq

\ni Machian Thin Sandwich 5) Another product of Machian Thin Sandwich 3) is that one can substitute the resultant extremizing $\mF^i$ back into the relational GR action. 
This can then be taken as an ab initio new action.

\ni Moves 2) to 5) constitute a reduction; together with 4.b), they render the Machian Thin Sandwich a subcase of Best Matching.  

\ni Machian Thin Sandwich 6) is that moves 3) and 4) also permit construction of the extrinsic curvature through the computational formula
\ni\beq
\mK_{ab} =  \frac{\pa_{\underline{\sF}} \mh_{ab}}{2 \, \pa \mI} \mbox{ }  .  
\label{K-Comp-A-2}
\eeq 
Sec \ref{Hyp-Kin} then considers yet further completion of the thin-sandwich prescription in terms of constructing the whole of the universal hypersurface kinematics \cite{Kuchar76II}.

\mbox{ }   

\ni See \cite{AHH} for cases of Thin Sandwiches which are simpler to solve, for all that the outcome of these is rather unexpected.  
Finally, see \cite{ABook} for restrictions on $\FrQ$, $\FrG$ pairings.  
Other pairings go awry within the present Subsec's premises on group theoretic grounds: no such group, or no natural group action of this on $\FrQ$.   
Some more cases go awry at the level of Constraint Closure, as per the below TRi version of Sec \ref{CC}.

\subsection{TRiPoD and Constraint Closure}\label{TRi-CC}

One can expand the manner in which Temporal and Configurational Relationalism stack to how to prevent going through any more gates causing one to get kicked out of the TR gate.
A first instalment of this is TRiPoD \cite{TRiPoD}. 
I.e. reformulating the Principles of Dynamics (PoD) to be TR incorporating -- TRiPoD -- ensures that all subsequent considerations of other Problem of Time facets 
within this new paradigm do not violate the TR initially imposed.
This is a homothetic tensor calculus  \cite{TRiPoD}, with changes $\d Q$ (or $\pa \mQ$ for field theories) assigned the unit covector weighting \cite{TRiPoD}.

The TRiPoD version of the PoD is similar in spirit to Dirac's introduction of multiple further notions of Hamiltonian to start to deal systematically with constrained systems, 
by appending diverse kinds of constraints with diverse kinds of Lagrange multipliers.
A differential ($\d$ for finite theories, $\pa$ for field theories) version of the Hamiltonian is required to be TRi.  
This affects such as the total Hamiltonian, which is now approached instead by appending cyclic differentials\footnote{To avoid confusion, 
note that `cyclic' in `cyclic differential' just means the same as `cyclic' in cyclic velocity, rather than implying some particular kind of differential itself. 
Thus nothing like `exact differential' or `cycle' in algebraic topology -- which in de Rham's case is tied to differentials -- is implied.} 
rather than Lagrange multipliers.

This gives rise to a variant of the Dirac approach based on a cyclic differential almost-Hamiltonian, 
and a matching variant of the Dirac Algorithm [with entities 2) and 4) now involving cyclic differentials rather than Lagrange multipliers].  
Using these (and TRi reformulation of constraint smearing) one can advance to Constraint Closure. 
This is in fact a sine qua non, for if a Constraint Closure Problem arises, 
it can invalidate one or both of the above Configurational and Temporal Relationalism implementations, or, in a more severe form, kill off the theory.
This is a big issue due to the `gates problem'. 
Appending is now done using cyclic differentials, e.g. $\pa \cH = \pa\mI \, \scH + \pa \mF^i \scM_i$ for GR's `differential almost' \cite{TRiPoD}\footnote{
In the Manifestly Reparametrization Invariant formulation, an `almost' Hamiltonian is involved since auxiliary variables' velocities remain. 
Then in passing to the geometrical formulation dual to the Manifest Parametrization Irrelevant formulation, auxiliary variables' differentials remain, 
so it is a `differential almost' Hamiltonian.}
`total Hamiltonian density' \cite{Dirac}.

\mbox{ } 

\ni Finally, the Dirac algebroid now takes the TRi presentation

\ni\beq
\mbox{\bf \{} ( \scM_i | \pa \mL^i ) \mbox{\bf ,} \, ( \scM_j | \pa \mM^j ) \mbox{\bf \}} = ( \scM_i | \, [ \pa \mL, \, \pa \mM ]^i )  \mbox{ } ,
\label{TRi-Mom,Mom}
\eeq

\ni \be
\mbox{\bf \{} ( \scH \, | \, \pa \mK ) \mbox{\bf ,} \,(    \scM_i  |    \pa \mL^i    ) \mbox{\bf \}} = (    \pounds_{\pa \underline{\sL}} \scH    |    \pa \mK    ) \mbox{ } , 
\label{TRi-Ham,Mom}
\ee

\ni \be 
\mbox{\bf \{} ( \scH  \, | \, \pa \mJ ) \mbox{\bf ,} \, (    \scH    |    \pa \mK    )\mbox{\bf \}}  = 
(   \scM_i    |    \pa \mJ \, \overleftrightarrow{\pa}^i \pa \mK    ) = (   \scM_i \mh^{ij}   |    \pa \mJ \, \overleftrightarrow{\pa}_j \pa \mK    ) \mbox{ } . 
\label{TRi-Ham,Ham}
\ee
\ni There are then two directions one can take (Fig \ref{Gates-2}.e). 
For now we have no preference as to in which order they are taken:it looks to be parallel processing, at least to the current level of understanding.

\subsection{TRi version of Expression in terms of Beables/Observables}\label{TRi-PoB}

This is one of the two directions.
This facet has to be considered after Constraint Closure because of basic but hitherto as far as I know totally overlooked fact:  
that beables algebraic structures only close if based upon closing subalgebraic structures of constraints \cite{ABeables}.
It is not different from the usual Expression in terms of Beables as regards practical mathematics, involving solving the same equations \cite{ABeables2} 
modulo formulation of accompanying smearing in field-theoretic case. 

\mbox{ } 

\ni In the field-theoretic case, $\pa$-smearing is induced by the constraints themselves being beables, albeit in a trivial sense.
I.e. the \K beables or observables condition is now

\ni\beq
\mbox{\bf \{} ( \scM_i | \pa \mL^i ) \mbox{\bf ,} \, ( \iK_{\sfK} | \pa \mV^{\sfK} ) \mbox{\bf \}} \mbox{ } `=' 0  \mbox{ } , 
\label{TRi-Mom,B}
\eeq
whereas Dirac beables or observables additionally obey 
\ni\beq
\mbox{\bf \{} ( \scH  \, |  \, \pa \mJ   ) \mbox{\bf ,} \, ( \iD_{\sfD}| \pa \mW^{\sfD} ) \mbox{\bf \}} \mbox{ } `=' 0  \mbox{ } .   
\label{TRi-Ham,B}
\eeq
See \cite{CapeTown} for a TRi version of model arenas for which classical such can be constructed.

\subsection{TRi version of Spacetime Construction}\label{TRi-SCP}

The other direction is to promote working with the TRi formulation of geometrodynamical GR to working with a family of theories \cite{RWR, AM13}. 
(RPMs \cite{BB82, B03, FileR, AMech} and minisuperspace \cite{AMini1} are arenas in which the Spacetime Construction and Foliation Independence steps do not occur 
and are largely trivial due to existence of hypersurfaces privileged homogeneity respectively.) 
In this way, the classical relational approach is already completed modulo the Dirac beables caveat posed in \cite{ABeables2}.
Thus look for model arenas which do exhibit these and more simply than full GR, e.g. 1 + 1 or 2 + 1 GR, slightly inhomogeneous cosmology \cite{AHH} or midisuperspaces such as Gowdy models.

\mbox{ } 

\ni Now one can start with a trial action which combines both Temporal and Configurational Relationalism:$^{\ref{MN-Ansatze}}$ 

\ni\be
\FS^{w,y,a,b} = \int \int_{\bigupsigma} \d^3x \sqrt{  \overline{a \, \mR + b}  } \, \pa \ms_{w,y} \mbox{ } . 
\label{trial} 
\ee
Then the TRi version of the Dirac Algorithm has

\ni\beq
\mbox{\bf \{} (    \scH_{x,y,a,b}   \, | \,   \pa \mJ    ) \mbox{\bf , } \, (    \scH_{x,y,a,b}   \, | \,   \pa \mK    ) \mbox{\bf \}}
\eeq
giving the TRi 2-form obstruction term

\ni\be
2 \times a \times y \times \{ 1 - x \} \times (    \mD_i \mp \, | \, \pa \mJ \, \overleftrightarrow{\pa}^i \pa \mK    ) \mbox{ } .  
\label{TRi-4-Factors}
\ee 
The accompanying `type of local relativity' matter results of Sec \ref{SCP} indeed carry over to the TRi case \cite{AM13}; 
see there and \cite{Lan2} for which other conjectures about matter made in \cite{RWR} hold and which have been disproven by counterexample.  

\mbox{ }

\ni The corresponding Spacetime Relationalism then works out as per Sec \ref{SRel}.

\section{Space-time split of GR}\label{Gdyn}

We next turn to setting up consideration of foliations. 
Moreover, it is important in a work about layers of foliation structure, and ultimately leading to work on layers of mathematical structure also, 
to work layer by layer in both of these senses.
The former post-dates ADM, though e.g. \cite{Gour} also exhibits some of this due to such layering also being a useful distinction to make in setting up Numerical Relativity calculations.
[These are geared toward such as black hole collisions and the dynamics of other expected sources of intense bursts of gravitational radiation.]

\subsection{Underlying topological manifold level structure}\label{Top-Sigma}
%

In conventional dynamical approaches to GR, preliminarily choosing a residual notion of space is required, 
in the sense of a 3-surface that is a fixed topological manifold $\bupSigma$ shared by {\sl all} the spatial configurations under consideration.  
I.e. dynamical GR approaches such as geometrodynamics are built subject to the restriction of not allowing for topology change: 
they are just `manifold topolostatics' rather than `manifold topolodynamics'.
This means that geometrodynamics covers a more restricted range of spacetimes than the spacetime formulation of GR does: those of spacetime topology $\bupSigma \times \bigtau$. 
In this Article, $\bupSigma$ is compact without boundary for simplicity, and, at least temporarily, on relational grounds.\footnote{Closed spaces have traditionally been regarded
as more Machian than open spaces, 
though the day will come in which unrestricted and dynamically evolving notions of space will be regarded as {\sl even more} background independent \cite{ASoS}.}
%
For now at least, we confine ourselves to a subset of GR's solution space so as to be able to study its dynamics within 
what mathematical methodology is currently known and accepted among physicists.  
Further restrictions are placed on topology in Sec \ref{Ev-Eq}. 
Finally, I use the notation $\bigupsigma$ in place of $\bupSigma$ if a 3-space is treated in isolation rather than as a slice within spacetime 
(in a sense made precise in Sec \ref{Single-Hyp}). 
For many purposes, one can also take a finite piece of space $\mS \subset \bupSigma$, rather than a whole space $\bupSigma$.

\subsection{Differential-geometric level structure}\label{DLS}

Assume additionally assumes that $\bigupsigma$ (or whichever of the preceding Sec's variants) carries differentiable manifold structure.  
The maps preserving this level of structure are diffeomorphisms, $Diff(\bigupsigma)$.

\subsection{Metric-level structure}\label{MLS}

$\langle \bupSigma, \mh_{ij} \rangle$ inherits spatiality from how it sits within $\langle \FrM, \mg_{\mu\nu} \rangle$. 
But to ensure that $\bigupsigma$ is indeed cast in a spatial role, this is directly equipped with a specifically Riemannian (positive-definite) 3-metric, $\bh$ with 
components $\mh_{ij}(\ux)$.

$\mbox{Riem}(\bupSigma)$ is then GR's configuration space consisting of all $\bh$'s on that particular fixed topological manifold $\bupSigma$.  
[denoted instead $\mbox{Riem}(\bigupsigma)$ if one does not wish to presuppose spacetime].  
The latter occurs e.g. in investigation of geometrodynamical theories in general,
rather than in treatment of specifically the geometrodynamics that is obtained by splitting GR spacetime and GR's Einstein field equations.

\subsection{Single-hypersurface concepts}\label{Single-Hyp}
%
{            \begin{figure}[ht]
\centering
\includegraphics[width=0.3\textwidth]{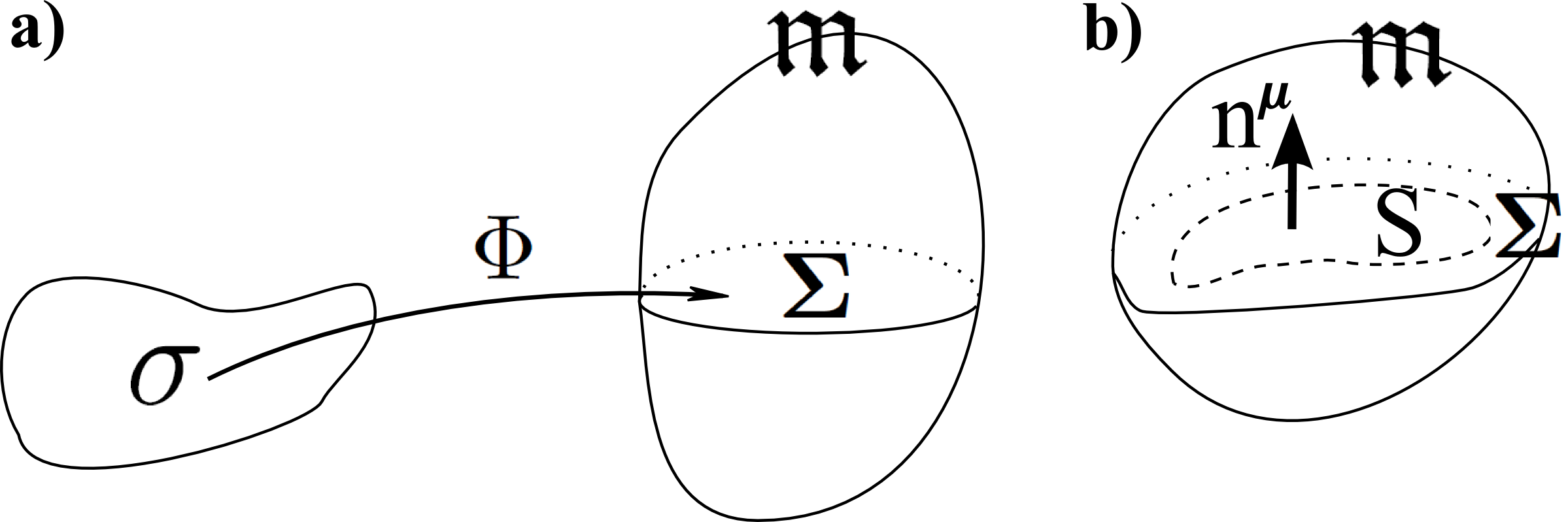}
\caption[Text der im Bilderverzeichnis auftaucht]{\footnotesize{a) Embedding $\Phi$ from 3-space $\bigupsigma$ to hypersurface $\bupSigma$ within spacetime $\FrM$.
b) The normal $\underline{\mn}$ to a hypersurface, and a local region $\mS \subset \bupSigma$.   } }
\label{Space-to-Hypersurface}\end{figure}            }

\ni Let us now consider passing \cite{Gour} from a 3-space $\bigupsigma$ to a spatial hypersurface $\bupSigma$ embedded in a spacetime $\FrM$ -- some `surrounding' space it `sits' in.   
More formally, a {\it hypersurface} $\bupSigma$ within $\FrM$ -- Fig \ref{Space-to-Hypersurface}.a) -- is the image of a plain spatial 3-manifold $\bigupsigma$ under a particular kind of 
map: an {\it embedding} $\bupPhi$.    
This construction can also be applied locally \cite{Wald}: embedding a piece  $\mbox{\Large $s$}$ of spatial 3-surface as a piece $\mS$ of hypersurface.  
Hypersurfaces are more generally characterized as surfaces $\FrHgen$ within a higher dimensional manifold $\FrMgen$ that are of {\it codimension 1}: 

\ni  dim($\FrMgen$) -- dim($\FrHgen$) = 1.

Next define the {\it normal}    $\mn^{\mu}$ to the hypersurface $\bupSigma$ (Fig \ref{Space-to-Hypersurface}.b) and 
                the {\it projector} $\mP^{\mu}\mbox{}_{\nu} := \delta^{\mu}\mbox{}_{\nu} + \mn^{\mu}\mn_{\nu}$  onto $\bupSigma$.  
The spacetime metric is furthermore said to {\it induce} the spatial metric on the hypersurface.
This induced metric is both an intrinsic metric tensor on space, $\mh_{ij}$ and a spacetime tensor $\mh_{\mu\nu}$. 
This exemplifies the current Article's second type of duality: that {\it hypersurface tensors} are both spatial and spacetime tensors.
Hypersurface tensors are defined as tensor such that for each `independent index' $\Theta_{\perp\nu...\omega} := \Theta_{\mu\nu...\omega}\mn^{\mu} = 0$.  
In the case of $\mh_{\mu\nu}$, since this is symmetric, $\mh_{\mu\nu}\mn^{\mu} = 0$ is a sufficient condition for this to be a hypersurface tensor, and this condition is indeed met. 
See Sec \ref{Hyp-Geom} for a further geometrical interpretation of the induced metric.
Finally note that once metric geometry becomes involved, it is isometric embeddings that one is dealing with.

Then {\it extrinsic curvature} of a hypersurface is the rate of change of the normal $\mn^{\mu}$ along a hypersurface,     

\ni \beq
\mK_{\mu\nu} := \mh_{\mu}\mbox{}^{\rho}\nabla_{\rho}\mn_{\nu} \mbox{ } .  
\label{K-def}
\eeq
This is a measure of the hypersurface's bending relative to its ambient space (i.e. spacetime in the case of geometrodynamics). 
Extrinsic curvature is symmetric and a hypersurface tensor; given the first property, $\mK_{\mu\nu}\mn^{\nu} = 0$ following from (\ref{K-def}) suffices to establish the second.
The above definition is then the primary one from the hypersurface-assumed side of the underlying duality; in this way, it antecedes the computational formula (\ref{K-Comp-ADM}).

Induced metric and extrinsic curvatures are `packaged together' by also being the {\it first} and {\it second fundamental forms} respectively, 
which between them contain the information about how a hypersurface is embedded in an ambient manifold.

Finally, Gauss' well known Theorema Egregium relating extrinsic and intrinsic curvature generalizes to\footnote{The (4) labels denote 
the corresponding spacetime versions of the objects thus indexed.} 

\ni \be
\mbox{(Gauss equation) }     \mbox{ }   \mR^{(4)}_{abcd} = \mR_{abcd} + 2 \mK_{a[c}\mK_{d]b} \mbox{ } , 
\label{Gauss-as-proj} 
\ee

\ni \be
\mbox{(Codazzi equation) }   \mbox{ }    \mR^{(4)}_{\perp abc} =  2 \mD_{[c} \mK_{b]a} \mbox{ } .
\label{Cod-as-proj}  
\ee
For now, these are to be viewed in terms of the left-hand sides being projections of the Riemann tensor which are then computed out to form the right-hand side.
%
%
%
\subsection{Two-hypersurface and foliation concepts}\label{Fol-Intro}

At least a thin one-sided infinitesimal neighbourhood of $\bupSigma$ (Fig \ref{Infinitesimal-Fol}.a) is required for a number of further notions \cite{Gour}.    
On the other hand, the notion of foliation (Fig \ref{Infinitesimal-Fol}.b) applies to an extended piece of spacetime, 
i.e. usually involving more 

\ni than just two infinitesimally-close slices.

Considering an infinitesimal limit of two neighbouring hypersurfaces, extrinsic curvature can furthermore be cast in the form of a Lie derivative, 

\ni \beq
\mK_{\mu\nu} = \pounds_{\underline{\sn}}\mh_{\mu\nu}/2 \mbox{ } .
\label{K-Lie}
\eeq
This observation offers immediate manifest proof of the aforementioned symmetry. 
%
{            \begin{figure}[ht]
\centering
\includegraphics[width=0.4\textwidth]{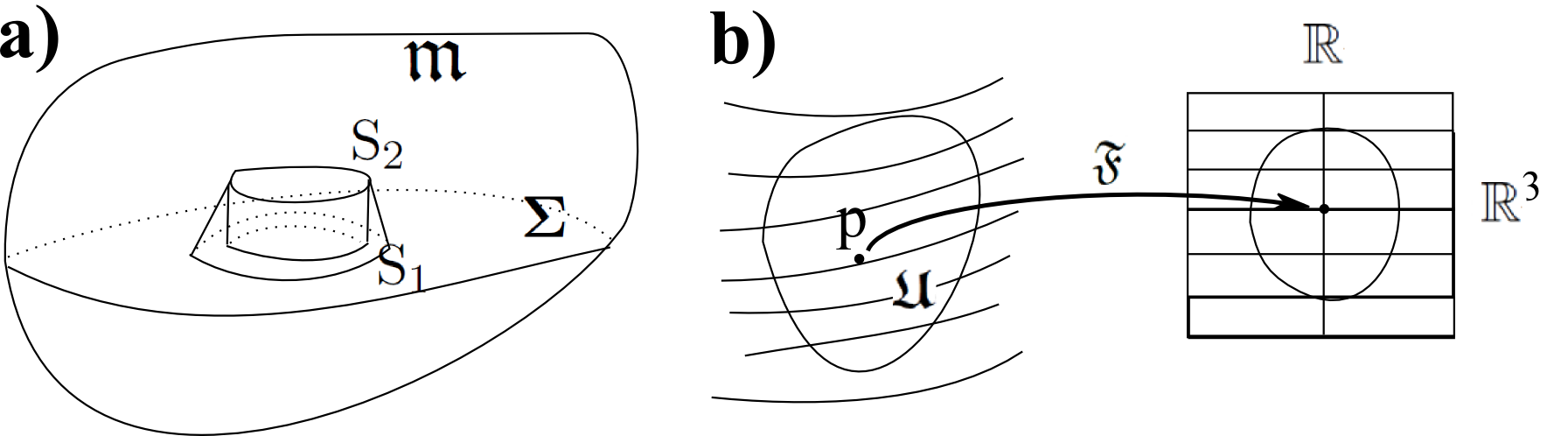}
\caption[Text der im Bilderverzeichnis auftaucht]{        \footnotesize{a) The set-up for $\mS$ a local in space piece of a spatial slice.  
b) Illustrating the nature of {\it foliation} $\FrF$: a `decorated' or rigged version of the more familiar definition of chart.}  }
\label{Infinitesimal-Fol}\end{figure}          }

\ni {\sl In GR, each foliation by spacelike hypersurfaces is to be interpreted in terms of a choice of time} $\mt$ {\sl and associated `time flow' vector field} $\mt^{\mu}$.  
There are an infinity of choices of such a $\mt$ (which e.g. \cite{Wald} terms  a `global time function').  
The spatial hypersurfaces correspond to constant values of that $\mt$.  

\mbox{ }  

\ni For Minkowski spacetime, $\mt$ and $\mt^{\mu}$ already exist as fully general entities, but they are usually chosen via a global inertial coordinate system \cite{Wald},  
and of course this ceases to exist in the general GR case.

For dynamical formulations of GR, one usually demands the spacetime to be time-orientable so that it is always possible to consistently allocate notions of past and future.

$\mt^{\mu}$ and $\mt$ are restricted by 

\ni\beq
\mt^{\mu}\nabla_{\mu} \mt =  1 \mbox{ and } \mbox{ }  \ms^{\mu}\nabla_{\mu} \mt =  0 
\label{restrict}
\eeq
for any tangential $\ms^{\mu}$.
Then if these hold, it is consistent to

\ni \beq
\mbox{identify } \mt^{\mu}\nabla_{\mu}  \mbox{ with } \pa/\pa \mt \mbox{ } \mbox{ and then}
\eeq

\ni \beq
\mbox{identify }  \pa/\pa \mt \mbox{ with }  \pounds_{\underline{\st}}  \mbox{ } ,
\label{pa-as-Lie}
\eeq
in the sense of (string of projectors) $\times \, \pounds_{\underline{\st}} \Theta$ for whichever tensor field $\Theta$.

\section{ADM split of spacetime metric}\label{ADM}
%
{            \begin{figure}[ht]
\centering
\includegraphics[width=0.3\textwidth]{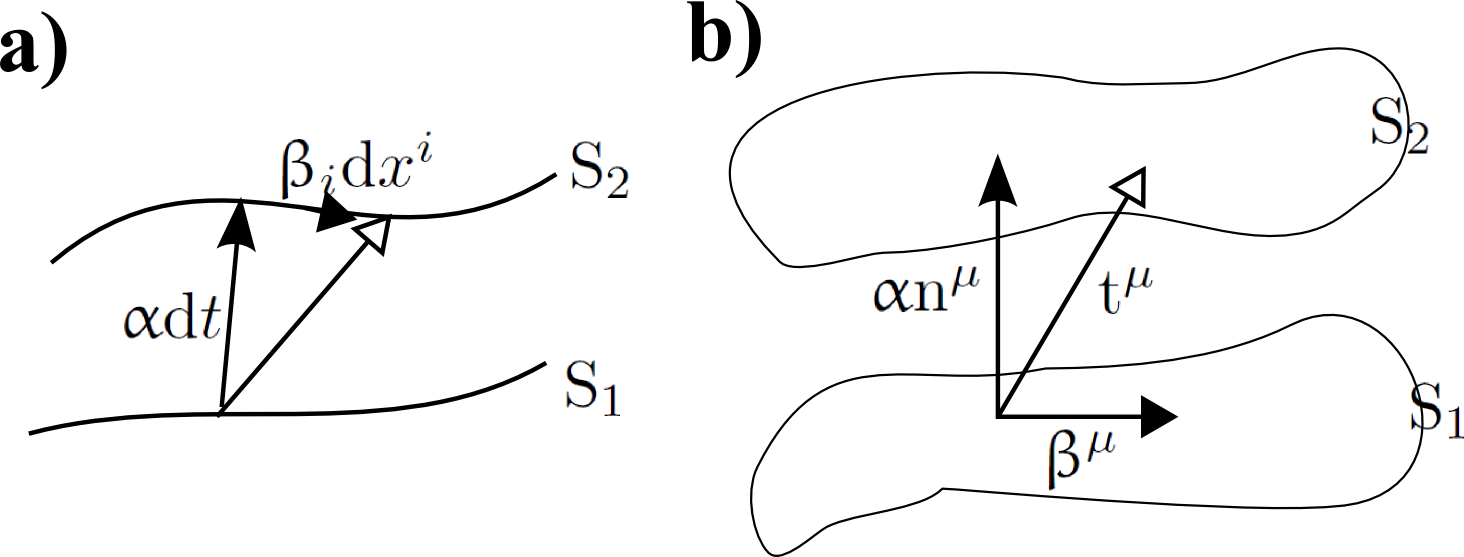}
\caption[Text der im Bilderverzeichnis auftaucht]{        \footnotesize{a) ADM 3 + 1 split of a region of spacetime, with lapse $\upalpha$ and shift $\upbeta^i$.
b) Local presentation of $\mt^{\mu}$, $\mn^{\mu}$, $\upbeta^{\mu}$ split.
The white arrows denote time flow alias deformation, the flat-backed black arrow denotes shift, and the slightly curved-bacjed arrow denotes normal.
}  }
\label{Infinitesimal-Fol-2}\end{figure}          }

ADM \cite{ADM} then split the spacetime metric into induced metric $\mh_{ij}$, shift $\upbeta^i$ and lapse $\upalpha$ pieces (Fig \ref{Infinitesimal-Fol-2}):

\ni \beq
\mg_{\mu\nu} =
\left(
\stackrel{    \mbox{$ \upbeta_{k}\upbeta^{k} - \upalpha^2$}    }{ \mbox{ }  \mbox{ }  \upbeta_{j}    } \stackrel{    \mbox{$\upbeta_{i}$}    }{  \mbox{ } \mbox{ }  \mh_{ij}    }
\right)
\mbox{ with inverse } \mbox{ } 
\mg^{\mu\nu} =
\left(
\stackrel{    \mbox{$ - {1}/{\upalpha^2}$}    }{ \mbox{ }  \mbox{ }   {\upbeta^j}/{\upalpha^2}    }
\stackrel{    \mbox{$   {\upbeta^i}/{\upalpha^2}$}    }{  \mbox{ } \mbox{ }  \mh^{ij}  - {\upbeta^i\upbeta^j}/{\upalpha^2}  }
\right)
\mbox{ } 
\label{ADM-split}
\eeq
and square-root of the determinant 
$ 
\sqrt{|\mg|} = \upalpha\sqrt{\mh}.
$
This split is often presented for a foliation, though 2 infinitesimally close hypersurfaces suffices.

In the ADM formulation, $\mt^{\mu}$ has the tangential--normal split:

\ni \be
\mt^{\mu} = \upbeta^{\mu} + \upalpha \, \mn^{\mu} \mbox{ } . 
\label{tbn}
\ee
This serves to define $\upbeta^{\mu} := \mh^{\mu\nu}\mt_{\nu}$ as the {\it shift} (displacement in identification of the spatial coordinates between 2 adjacent slices, 
which is an example of {\it point identification map} \cite{Stewart}). 
Note that $\upbeta^{\mu}$ are spacetime components, but of a hypersurface vector $\vec{\upbeta}$ since $\mn_{\mu}\mh^{\mu\nu} \mt_{\nu} = 0$, 
so this object can be denoted $\upbeta^i$ also; Kucha\v{r}'s hypersurface approach prefers the $\vec{\upbeta}$ form.  
Additionally, $\upalpha := - \mn_{\gamma}\mt^{\gamma}$ is the {\it lapse} (`time elapsed'), which may be interpreted as duration of proper time $\d\uptau = \upalpha(t, x^i)\d t$. 
In a simpler setting \cite{MTW} lapse $\upalpha$  is $\pa \mT/\pa t = \pa \mT/\pa t$, i.e. d(proper time)/d(coordinate time), 
                            and shift $\upbeta^i$ is $\pa \mX^i/\pa t$,             i.e. d(difference in coordinate grid)/d(coordinate time).

Now the normal is $\mn^{\mu} = [1, - \uupbeta]/\upalpha$,   
and a computational form for the extrinsic curvature is 

\ni \beq
\mK_{ij} =  \frac{  \dot{\mh}_{ij} - \pounds_{\vec{\upbeta}}\mh_{ij}  }{  2 \, \upalpha  } =: \frac{  \Circ_{\vec{\upbeta}}{\mh}_{ij}  }{  2 \, \upalpha  }  \mbox{ } .   
\label{K-Comp-ADM-2}
\eeq

\section{A-split of spacetime metric}\label{A}
%
{            \begin{figure}[ht]
\centering
\includegraphics[width=0.3\textwidth]{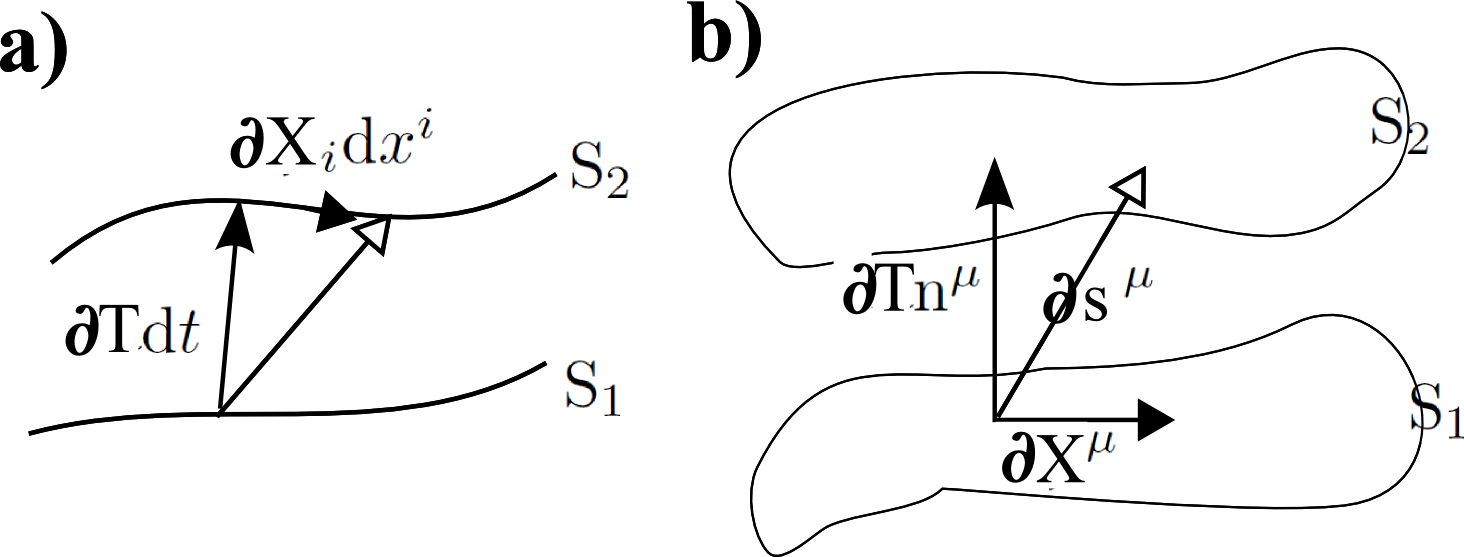}
\caption[Text der im Bilderverzeichnis auftaucht]{        \footnotesize{a) A 3 + 1 split of a region of spacetime, with lapse $\upalpha$ and shift $\upbeta^i$.
b) Local presentation of $\pa \ms^{\mu}$, $\mn^{\mu}$, $\upbeta^{\mu}$ split.]}  }
\label{TRi-Infinitesimal-Fol}\end{figure}          }
 
\ni Supposing however that one arrives at spacetime from relational first principles via the almost-Dirac Algorithm's spacetime construction.
Then it is instead natural to think of the kinematics of spacetime splits in the following alternative terms.  

\mbox{ }

\ni \underline{Highlight 1}. Start\footnote{This is a tentative starting-point, which then points to Sec \ref{Fol-Intro} needing re-evaluation as well.}
by splitting the spacetime metric into induced metric $\mh_{ij}$, partial differential of the frame $\pa \mX^i$ and partial differential of the instant $\pa \mT$ pieces 
(Fig \ref{TRi-Infinitesimal-Fol}):

\ni \beq
\mg_{\mu\nu} =
\left(
\stackrel{    \mbox{$ \pa \mX_{k}\pa \mX^{k} - \pa \mT^2$}    }{ \mbox{ }  \mbox{ }  \pa\mX_{j}    } 
\stackrel{  \mbox{ }  \mbox{ }  \mbox{$\pa\mX_{i}$}    }{  \mbox{ } \mbox{ } \mbox{ }  \mh_{ij}    }
\right)
\mbox{ } \mbox{ with inverse } \mbox{ } 
\mg^{\mu\nu} =
\left(
\stackrel{    \mbox{$ - {1}/{\pa \mT^2}$}    }{ \mbox{ }  \mbox{ }   {\pa\mX^j}/{\pa\mT^2}    }
\stackrel{    \mbox{$  \mbox{ } {\pa\mX^i}/{\pa\mT^2}$}    }{  \mbox{ } \mbox{ } \mbox{ } \mbox{ } \mh^{ij}  - {\pa\mX^i\pa\mX^j}/{\pa\mT^2}  }
\right) \mbox{ } 
\label{A-split}
\eeq
The form of the metric in turn means that for the spacetime interval to be Tri-invariant, then $[\d t, \d x^i]$ is of the form $[\pa^{-1} u, y^i]$ as regards TRi-rank. 
I.e. unsurprisingly, the spatial part is a TRi-scalar, whereas the GR coordinate time element $\d t$ is a TRi-{\sl vector}.\footnote{Since this is a secondary entity 
in the Relational Approach, it does not get to dictate the notation, which has been chosen to match up rather for change $\d Q^{\sfA}$ or $\pa \mh_{ij}$ being a covector.
Consequently the emergent time element $\d t^{\te\tm(\tc\tl)}$ or $\pa \mt^{\te\tm(\tc\tl)}$ is also a covector. 
GR coordinate time then so happens to scale in oppositely to this.}
%
The corresponding split of the inverse metric has three opposing-weight TRi pieces: a 2-tensor, a vector and a scalar.
The square-root of the determinant itself is a TRi-covector

\ni \be 
\sqrt{|\mg|} = \pa \mT \sqrt{\mh} =: \pa \Delta \mbox{ }.
\ee
\ni \underline{Highlight 0} In considering the analogue of $\mt^{\mu}$ in the A formulation, 
it is noticed that the time flow vector field becomes a TRi-covector $\pa\ms^{\mu}$. 
Thus TRi-foliations are interpreted in terms of a choice of time $\mt$ and an associated time flow vector field TRi-covector $\pa \ms^{\mu}$. 
Also then decomposition (\ref{tbn}) needs to be replaced by 

\ni \be
\pa \ms^{\mu} = \pa\mX^{\mu} + \pa\mT \, \mn^{\mu} \mbox{ } . 
\label{sXn}
\ee

\ni \underline{Highlight 2} Then in place of (\ref{restrict}), $\pa \ms^{\mu}$ and $\mt$ are restricted by 

\ni\beq
\pa \ms^{\mu}\nabla_{\mu}\mt = 1 \mbox{ and } \mbox{ } \pa \mw^{\mu}\nabla_{\mu}\mt = 0
\eeq
for any tangential $\pa \mw^{\mu}$; formulated in this way, entities decompose into time flow and tangential parts in a TRi-covariant manner.  
Then if these hold, it is consistent to identify $\pa \ms^{\mu}\nabla_{\mu}$ with both $\pa/\pa \mt$ and with $\pounds_{\pa \underline{\ms}}$.
On the other hand, $\mn^{\mu}$ remains a TRi-scalar.

\mbox{ } 

\ni \underline{Highlight 3} Indeed, rearranging (\ref{sXn}) to make $\mn^{\mu}$ the subject,

\ni \beq
\mn^{\mu} = [\pa \mn^0, - \pa \mX^i]/\pa \mT
\eeq
and $\pa \ms^0$ is numerically 1 but remains none the less a TRi-covector. 
With $\mn^{\mu}$ being a TRi-scalar, sa are the projector ${\mP^{\mu}}_{\nu}$, the notion of hypersurface tensor, and the extrinsic curvature $\mK_{\mu\nu}$.

These logically prior statuses of $\mn^{\mu}$ and $\pa \ms^{\mu}$ deduced, one can return and check that the equations which defined lapse and shift carry over to their TRi counterparts.

\mbox{ } 

\ni \underline{Highlight 4} Indeed then the {\it partial differential of the frame} is $\pa \mX^{\mu} = \mh^{\mu\nu}\pa \ms_{\nu}$: 
another formulation of the above point identification map.
This is also a hypersurface tensor by $\mn_{\mu} \pa \mX^{\mu} = \mn_{\mu} \mh^{\mu\nu} \pa \ms_{\nu} = 0$, so it can be written $\pa \vec{\mX}$ or $\pa \underline{\mX}$, 
but the 3-space approach favours $\pa \vec{\mF}$ or $\pa \underline{\mF}$.  

\ni \underline{Highlight 5} Also, the {\it partial differential of the instant} is $-\mn_{\mu} \pa \ms^{\mu} = \pa \mT$: 
another formulation of duration of proper time, now in the furtherly simplified form of identifying $\uptau$ with $\mT(t, x^i)$ itself. 

\ni \underline{Highlight 6}. In this case, an explicit computational formula for the extrinsic curvature is

\ni \beq
\mK_{ij} =  \frac{ \pa{\mh}_{ij} - \pounds_{\pa \underline{\sX}}{\mh}_{ij}}{2 \, \pa\mT} =: \frac{ \pa_{\underline{\sX}}   \mh_{ij} }{  2 \, \pa\mT  }  \mbox{ } .     
\label{K-Comp-A}
\eeq
This is now in terms of the Best Matching derivative \cite{RWR}, whose general form is in terms of a Lie derivative correction on account of the shuffling procedure entailed.  
By the metric Lie derivative--Killing form identity, and under the change from A split to ADM split quantities,\footnote{To help keep track throughout this Article, 
the full diversity of notation in use for geometrodynamical auxiliaries is as follows.
ADM-and-BSW shift                    $\upbeta^i$, 
ADM presupposed lapse                $\upalpha$, 
BSW emergent lapse                   $\mN$, 
relational 3-space spatial frame     $\mF^i$, 
A-split of spacetime's spatial frame $\mX^i$, 
relational 3-space emergent instant  $\mI$, 
A-split of spacetime's instant       $\mT$.
The last two of these are numerically equal to proper time, $\tau$.  
The corresponding smearing variables are  $\upxi^i$   and $\upchi^i$, 
                                          $\upzeta$   and $\upomega$, 
										  unused, 
										  $\pa \mL^i$ and $\pa \mM^i$, 
										  $\pa \mY^i$ and $\pa \mZ^i$, 
										  $\pa \mJ$   and $\pa \mK$, 
									 and  unused. 
									 \label{Foo-Nota}}  
one can see that this is mathematically the same as (\ref{K-Comp-ADM}).
This is underlied by Barbour's Best-Matching derivative (\ref{BM-Der}) being the 3-space dual interpretation of Kucha\v{r}'s hypersurface derivative (\ref{Hyp-Der}).  
Moreover, (\ref{K-Comp-A})'s interpretation is now as {\it comparison of each geometrodynamical change with the STLRC}.  

\mbox{ }

\ni On the other hand, the Gauss and Codazzi equations are TRi-scalars, and so carry over.

\section{Further basic considerations of foliations}\label{Fol-2}

\subsection{Interpretation of foliations in terms of families of possible observers}\label{Obs}
%
{            \begin{figure}[ht]
\centering
\includegraphics[width=0.55\textwidth]{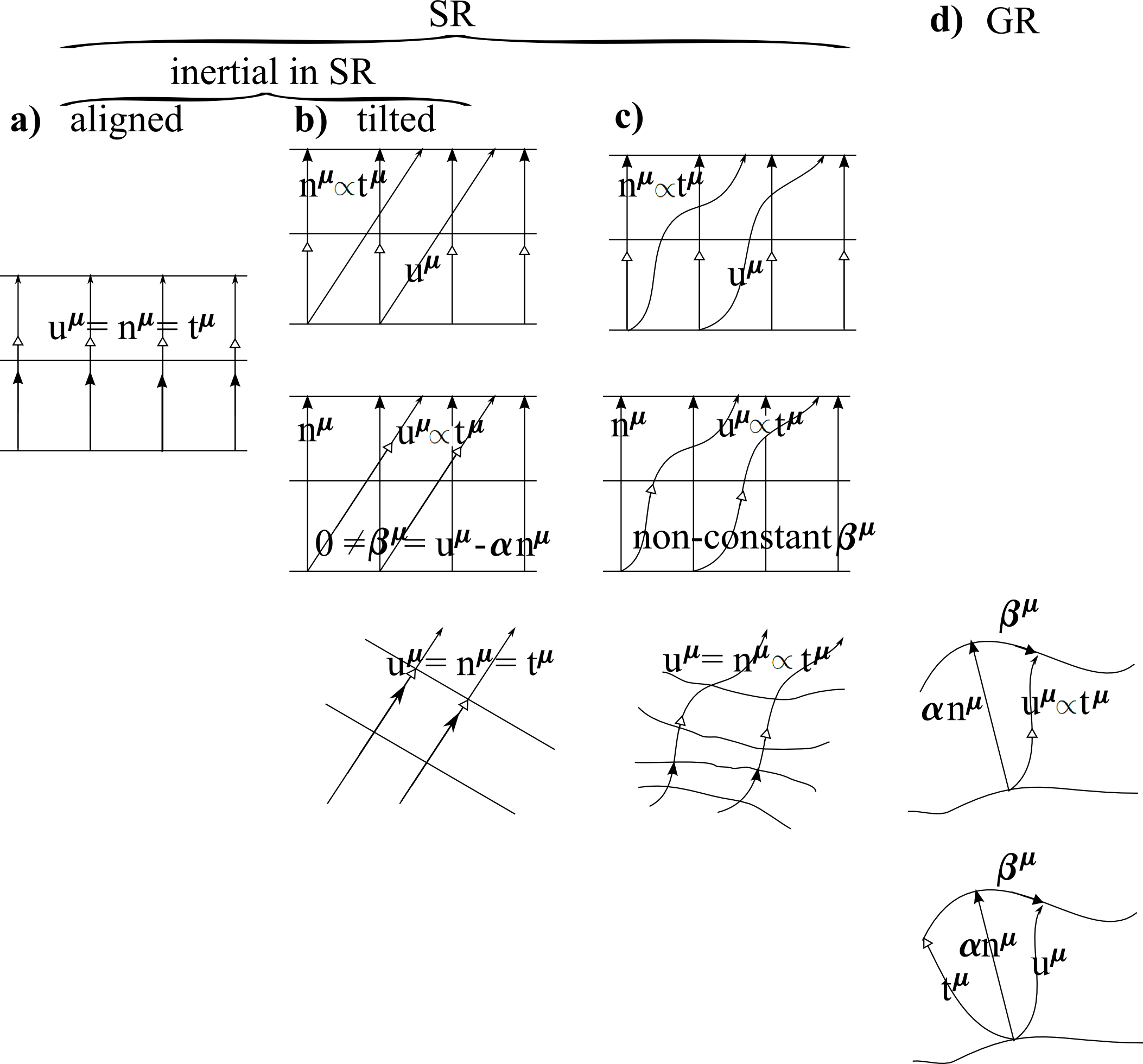}
\caption[Text der im Bilderverzeichnis auftaucht]{        \footnotesize{\ni a) Material flow $\muu^{\mu}$, normal vector field $\mn^{\mu}$ and time flow $\mt^{\mu}$ 
all coincide for Eulerian observers (depicted here in the SR case).

\mbox{ } 

\ni \underline{Highlight 7} is that the TRi version of this definition being meaningful at the level of homothetic tensor calculus 
requires multiplying $\muu^{\mu}$ and $\mn^{\mu}$ by the unit-valued TRi-covector.  

\mbox{ } 

\ni b) In this case, material flow is {\it tilted away} from the normals.  
Then two simple cases involve time flow aligned with each of these in turn.  
Now in some circumstances, the hypersurfaces to which the normal vector fields coincide have no physical significance. 
In the depicted SR case with homogeneous but tilted material flow, the third subfigure corresponds to dropping the initial inertial frame for 
a distinct material flow aligned inertial frame (rest frame).
However, in the isotropic cosmology analogue, there are hypersurfaces privileged by homogeneity, for which the preceding flexibility of changing to an equally simple frame is lost.
%

\ni c) Furthermore, if the material flow is not homogeneous, passing from the second row to the third amounts to leaving the inertial frames for a more complicated general frame.
It is in this context that even Minkowski spacetime has need of hypersurface deformations and indeed manifests the Dirac algebroid \cite{Dirac}.
Observers comoving with the material flow is quite a natural situation, 
though local physics is capable of distorting that with such as local Keplerian motion or `peculiar velocities' more generally. 

\ni d) In generic GR, there are no flat hypersurfaces to begin with. 
The most general situation for a family of observers (already existing for SR, though not depicted there) is for each to come in a `rocket' that is capable of arbitrary accelerations.  
The motion of this need not then be aligned with the normal or the material flow vector. 
Thus we have descended from privileged flat foliation of flat spacetime with double alignment to a generic hypersurface in a generic spacetime with no alignments. }  }
\label{Observer-Congruences}\end{figure}          }

\ni Each normalized $\mt^{\mu}$ [$\mt^{\mu}/||\underline{\mt}|| = \{\upalpha \, \mn^{\mu} + \vec{\upbeta}^{\mu}\}/          \sqrt{\upalpha^2 - \upbeta^2}$ 
$= 
                                                            \{\upalpha \, \mn^{\mu} + \vec{\upbeta}^{\mu}/\upalpha \}/\sqrt{   1 - \{\upbeta/\upalpha\}^2   } :=
                                                            \{\upalpha \, \mn^{\mu} + \vec{\mv}^{\mu} /   \upalpha\}/\sqrt{1 - \mv^2} =
															\upgamma [1, \underline{\mv}]$ for $\underline{\mv} := \underline{\upbeta}/\upalpha$ in the ADM parlance,
represents a distinct possible motion of a cloud of observers.
In the TRi case, $\underline{\mv} = \pa\underline{\mX}/\pa\mT$ and $\upgamma$ is a function of $\underline{v}$ alone.
Thus $\pa\ms^{\mu}/||\pa\underline{\ms}| = \upgamma [1, \underline{\mv}]$ is a TRi-scalar expression. 
In whichever of the two cases, then, these observers are held to be combing out spacetime rather than travelling on mutually-intersecting worldlines.
Elsewise, they have freedom of motion: `rocket engines' permitting each to accelerate independently of the others.

Foliations can be thought of as the as level surfaces of the scalar field notion of time, $\mt$.  
Here $\mt$ is taken to be smooth, with a gradient that is nonzero everywhere, ensuring that these level surfaces do not intersect anywhere.

\mbox{ }  

\ni We shall see in Sec \ref{Bub-Finger} that GR's well-known notion of {\it many-fingered time}, 
and the ray--wavefront dual concept (the current Article's third distinct type of duality) of {\it bubble time}, are concepts along the above lines.

\subsection{Completion of the curvature projection equations with further interpretations}\label{Ricci}

This Sec continues Sec \ref{Fol-Intro} by being 2-hypersurface or foliation concepts, but now involving the ADM or A split. 
The {\it Ricci equation} is the final projection of the Riemann tensor.
In the ADM formulation, 

\ni \be
\mR^{(4)}_{\perp a\perp b} =  \frac{\Circ_{\vec{\upbeta}} \mK_{ab} + \mD_b\mD_a\upalpha}{\upalpha} + {\mK_a}^c\mK_{cb} \mbox{ } .
\label{Ricci-as-proj}
\ee
\underline{Highlight 8}. On the other hand, in the A formulation, 

\ni \be
\mR^{(4)}_{\perp a\perp b} =  \frac{\pa_{\underline{\sX}} \mK_{ab} + \mD_b\mD_a\pa\mT}{\pa\mT} + {\mK_a}^c\mK_{cb} \mbox{ } .
\label{TRi-Ricci-as-proj}
\ee
Note how unlike the Gauss and Codazzi equations, the Ricci equation is specifically an at least infinitesimal foliation concept: 
whilst $\mK_{ab}$ belongs to a surface, $\dot{\mK}_{ab}$ or $\pa{\mK}_{ab}$ requires an infinitesimal foliation.

Three possible conceptually-distinct interpretations of the Gauss, Codazzi and Ricci equations are then as follows.

\mbox{ }

\ni 1) Top-down. 
Given a higher-$d$ manifold containing a hypersurface, how do its curvature components project onto that hypersurface 
(as a combination of intrinsic and extrinsic curvatures of that hypersurface)?

\ni 2) Down-up. 
Given a 3-surface's intrinsic geometry and how it is bent within its ambient 4-manifold, what can be said about the ambient manifold's intrinsic curvature?  

\ni 3) Intrinsic to extrinsic.  
Given the intrinsic geometry of both a 3-surface and of a 4-manifold, is there a bending by which this 3-surface fits within the 4-manifold as a hypersurface?

\mbox{ }  

\ni I.e. 1) constructs the geometry of a hypersurface within a given manifold. 
2) constructs the manifold locally surrounding the hypersurface. 
3) determines whether a given surface can be realized as a hypersurface within an also-given manifold.

\subsection{Space-time splits of the GR action}\label{Action-Splits}

\ni Under the ADM split, the Einstein--Hilbert action is recast as  (\ref{S-ADM})
This is obtained by {\sl decomposing} $\mR^{(4)}$ using a combination of contractions of the Gauss and Ricci equations 
and discarding a total divergence since $\bupSigma$ is without boundary.

The result of varying with respect to this action can be recognized as 3 projections of the spacetime Einstein tensor.
I.e. various combinations of contractions of the above Gauss, Codazzi and Ricci equations viewed as projection equations.
These equations can also be obtained by projecting the Einstein field equations themselves.

However many applications require adherence to the Principles of Dynamics in laying these out.    
Begin by considering the manifestly-Lagrangian form of the action, i.e. in terms of configurations and velocities, which are here $\mh_{ij}$ and $\dot{\mh}_{ij}$.

\mbox{ } 

\ni \underline{Highlight 9}. The A counterpart of this is$^{\ref{Foo-Nota}}$

\ni\beq
\FS_{\sA} = \int \d t \int_{\Sigma}\d^3x \, \pa \cK_{\sA} = \int \d t \int_{\Sigma}\d^3x\sqrt{\mh} \, \pa\mT \{\mK_{ab}\mK^{ab} - \mK^2   +  \mR\} \mbox{ } . 
\label{S-A}
\eeq
In this case, one is considering the manifestly-Jacobian form of the action, in terms of the Machian variables of configuration and change:  $\mh_{ij}$ and $\d \mh_{ij}$.
$\pa \cK_{\sA}$ is here the Lagrangian TRi-covector.

\subsection{The GR action endows $\mbox{Riem}(\bupSigma)$ with a metric geometry}\label{DeWitt-supermetric}

I next reformulate this action in terms of the configuration space geometry for GR.
(\ref{S-ADM})'s kinetic term contains the GR kinetic metric contracted into $\mK_{ab}\mK_{cd}$ and thus into $\dot{\mh}_{ab}\dot{\mh}_{cd}$.  
Moreover DeWitt's \cite{DeWitt67} 2-index to 1-index map $\mh_{ij} \mapsto \mh^A$ casts it in the standard form of two downstairs indices: $\mM^{abcd} \mapsto \mM_{AB}$.
Then $\overline{\cT}$ takes the form $\mM_{AB}\Circ_{\vec{\upbeta}}{\mh}^A\Circ_{\vec{\upbeta}}{\mh}^B$.  
[The capital Latin indices in this context run from 1 to 6.]
Pointwise, this is a --+++++ metric, and so, overall it is an infinite-dimensional version of a semi-Riemannian metric.  
The above indefiniteness is associated with the expansion of the universe giving a negative contribution to the GR kinetic energy.  
This is entirely unrelated to the indefiniteness of SR and GR spacetimes themselves.  
DeWitt additionally studied the more detailed nature of this geometry in \cite{DeWitt67}.

$\mM_{\sfA\sfB}$ is a TRi-scalar.
Thus substituting it into the ADM and A actions generates no new differences, producing the geometrical DeWitt form of the manifestly Lagrangian form of the ADM action

\ni \beq
\FS_{\sA\sD\sM-\sL\sD}  =  \int\d t\int_{\sbSigma}\d^3x \, \cL_{\sA\sD\sM-\sJ\sD} = \int\d t\int_{\sbSigma}\d^3x \, \upalpha \big\{\overline{\cT}_{\sA\sD\sM-\sL\sD}/\upalpha^2 +  
\overline{\mR}   - 2 \, \overline{\slLambda} \big\} \mbox{ } , \mbox{ } \mbox{ for } 
\overline{\cT}_{\sA\sD\sM-\sL\sD} = ||\Circ_{\vec{\upbeta}}\bh||_{\mbox{\scriptsize ${\bM}$}}\mbox{}^2/4 \mbox{ } .  
\label{S-ADM-LD}
\eeq
\underline{Highlight 10}. The geometrical DeWitt form of the manifestly Lagrangian A action is 

\ni \beq
\FS_{\sA-\sJ\sD} =  \int\d t\int_{\sbSigma}\d^3x \, \pa \cK_{\sA-\sJ\sD} =  
\int\d t\int_{\sbSigma}\d^3x\sqrt{\mh}\,\pa\mT \big\{ \overline{\cT}_{\sA-\sJ\sD} + \overline{\mR} - 2 \, \overline{\slLambda} \big\} 
\mbox{ } , \mbox{ } \mbox{ for } \overline{\cT}_{\sA-\sJ\sD} = ||\pa_{\underline{\sX}}\bh||_{\mbox{\scriptsize ${\bM}$}}\mbox{}^2/4 \mbox{ } .  
\label{S-A-JD}
\eeq

\subsection{GR's momenta}\label{GR-mom}

Now as well as extrinsic curvature being an important characterizer of hypersurfaces, it is relevant due to its bearing close relation to the GR momenta, whose ADM form is 

\ni \beq
\mp^{ij} :=  \frac{\delta \cL_{\sA\sD\sM}}{\delta \dot{\mh}_{ij}}
          = -\sqrt{\mh}\{ \mK^{ij} - \mK \mh^{ij} \}                       = \mM^{ijkl}\frac{\Circ_{\vec{\upbeta}}\mh_{ij}}{2 \, \upalpha} \mbox{ } .  
\label{Gdyn-momenta}
\eeq
I.e. the gravitational momenta are a densitized version of $\mK_{ab}$ modulo a trace term.  
%
%
Taking the trace, 

\ni \beq
\mp = - 2 \, \sqrt{\mh} \mK \mbox{ } . 
\label{Tr}
\eeq
\ni \underline{Highlight 11} On the other hand, TRiPoD necessitates a new definition of momentum, which applied to the A form of 

\ni the GR momenta gives

\ni \beq
\mp^{ij} := \frac{\delta \, \pa \cK_{\sA}}{\delta \, \pa {\mh}_{ij}} = \mM^{ijkl}\frac{\pa_{\underline{\sX}}\mh_{ij}}{2 \, \pa \mT} \mbox{ } , 
\label{Gdyn-TRi-momenta}
\eeq
with the equality and the subsequent trace formula (\ref{Tr}) remaining unaffected as TRi-scalar entities.

\subsection{GR's constraints}\label{GR-cons-and-sp-diff}

$\scH$ and $\scM_i$ are TRi-scalars.  
Being a TRi scalar, $\scM_i$'s interpretation is regardless of the formulation that begot it.
Namely, as its physical content residing not in the 3 degrees of freedom per space point (dofpsp) choice of point-identification 
but rather solely in terms of the 3-metric's {\sl other} 3 dofpsp, termed the {\it 3-geometry}. 
This is how GR is, more closely, a dynamics of 3-geometries (the diffeomorphism invariant information in the 3-metric) \cite{Battelle, DeWitt67} on the quotient configuration space 

\ni \beq
\mbox{Superspace($\bupSigma)$ := $\mbox{Riem}(\bupSigma)/Diff(\bupSigma)$} \mbox{ } . 
\eeq
\mbox{ } \mbox{ } However, interpreting the GR Hamiltonian constraint is tougher. 
It has a `purely-quadratic in the momenta' form, meaning it is a quadratic form plus a zero-order piece but with no linear piece.
As per Sec \ref{TR}, this property leads to the Frozen Formalism Facet of the Problem of Time.  

\mbox{ } 

\ni Also note that in terms of $\mK_{\mu\nu}$ (and setting $\slLambda = 0$), the constraints are  \cite{Darmois}

\ni \beq
0 = - \scH    = \mK^2 - \mK_{ij} \mK^{ij}  + \mR       = 2 \, \mG^{(4)}_{\perp\perp} \mbox{ } ,   
\eeq

\ni \be
0 = \scM_{i}  = 2\{\mD_{j}{\mK^{j}}_{i} - \mD_{i}\mK\} =  2 \, \mG^{(4)}_{i\perp} \mbox{ } .
\label{Mom}
\eeq
As indicated, the $\mK_{ij}$ forms of these constraints serve to identify \cite{Wald} 
these as contractions of the Gauss--Codazzi equations for the embedding of spatial 3-slice into spacetime: the `{\it Constraint--Embedding Theorem of GR}'.

\subsection{GR's evolution equations}\label{Ev-Eq}

$\bupSigma \times \biguptau$ is usually assumed to ensure good causal behaviour; moreover, this then prevents consideration of topology change. 
We also assume $\bupSigma$ is everywhere spacelike and a Cauchy surface for $\FrM$, by whose existence $\FrM$ is globally hyperbolic \cite{Wald}.  

\mbox{ } 

%
\ni In terms of the gravitational momenta, the ($\slLambda = 0$) ADM evolution equations are

\ni \beq
\Circ_{\vec{\upbeta}}\mp^{ij} = \sqrt{\mh} \left\{ \frac{\mR}{2} \mh^{ij}  -  \, \mR^{ij} + \mD^j\mD^i - \mh^{ij} \mD^2        \right\}  \upalpha  
                               - \frac{2\upalpha}{\sqrt{\mh}}            \left\{  \mp^{ic}{\mp_c}^j - \frac{\mp}{2} \, \mp^{ij} \right\} 
							   + \frac{2\upalpha}{\sqrt{\mh}}  \mh^{ij}  \left\{  \mp_{ij}\mp^{ij}  - \frac{\mp^2}{2}           \right\}             \mbox{ } .  
\label{ADM-Evol}
\ee
\underline{Highlight 12}. The corresponding A evolution equations are 

\ni \beq
\frac{\pa_{\underline{\sX}}\mp^{ij}}{\pa\mT} = \frac{\sqrt{\mh} \{ \mR \, \mh^{ij}/2  -  \, \mR^{ij} + \mD^j\mD^i - \mh^{ij} \mD^2 \}\pa\mT}{\pa \mT} 
                                 - \frac{2}{\sqrt{\mh}}         \left\{ \mp^{ic}{\mp_c}^j - \frac{\mp}{2} \mp^{ij} \right\}
							     + \frac{2}{\sqrt{\mh}} \mh^{ij}\left\{\mp_{ij}\mp^{ij} - \frac{\mp^2}{2}\right\}   \mbox{ } . 
\label{A-Evol}
\ee
In this case, recasting them in terms of extrinsic curvature does not keep one within a TRi-scalar form due to their higher-derivative multi-slice status. 
The ADM version of this is 

\ni \be
\frac{- \{\Circ_{\vec{\upbeta}} \mK_{ab} - \mh_{ab}\Circ_{\vec{\upbeta}}\mK\} - \mD_b\mD_a\upalpha + \mh_{ab}\mD^2\upalpha}{\upalpha}
- \left\{
2{\mK_a}^c\mK_{bc} - \mK\mK_{ab} + \frac{\mK_{ij} \mK^{ij} + \mK^2}{2}\mh_{ab}
\right\}
+ \mG_{ab} = \mG^{(4)}_{ab} = 0 \mbox{ } . 
\label{ADM-Evol-2}
\ee
which form completes the constraint equations as regards forming the remaining projection of ${\mG}^{(4)}_{\mu\nu}$,

\mbox{ } 

\ni \underline{Highlight 13}. The A counterpart is then that  

\ni \be
\frac{- \{\pa_{\underline{\sX}} \mK_{ab} - \mh_{ab}\pa_{\underline{\sX}}\mK\} - \{\mD_b\mD_a - \mh_{ab}\mD^2\}\pa\mT}{\pa\mT} - 
\left\{
2{\mK_a}^c\mK_{bc} - \mK\mK_{ab} + \frac{\mK_{ij} \mK^{ij} + \mK^2}{2}\mh_{ab}
\right\}
+ \mG_{ab} = \mG^{(4)}_{ab} = 0 \mbox{ } . 
\label{A-Evol-2}
\ee
The three ADM-split Einstein field equations can all be interpreted in terms of contractions of the Gauss--Codazzi--Ricci embedding equations, as can the three A-split ones.
In this way, the Constraint--Embedding Theorem to the {\it Constraint--Evolution--Embedding Theorem of GR}, 
which now has fine distinction between an ADM form and an A form through the last piece -- the GR evolution equations -- not being already-TRi.

\section{Approaches giving foliations a more primary status}\label{Fol-Theory}

\subsection{Single-slice concepts. i. topological and differentiable manifold levels}\label{Hyp-Geom}

This further develops firstly Sec \ref{Single-Hyp}'s consideration of a single hypersurface $\bupSigma$ within a manifold $\FrM$, 
from both the top-down and down-up routes outlined in Sec \ref{Ricci}.  
%
%
The idea now is to give a more global approach \cite{Bubble, Kuchar76I, Kuchar76II, Kuchar76III, Kuchar81}, 

\ni as opposed to how previous work \cite{ADM, Dirac} depends on choosing coordinates, which at best holds locally in general.

\mbox{ } 

\ni At the topological level, consider $\FrM = \bupSigma \times \FrI$ for some interval $\FrI$ as per Sec \ref{Top-Sigma}.  
Then the {\it Slice} operation involves identifying a particular slice $\bupSigma$ in $\FrM$. 
On the other hand, the {\it Project} operation involves keeping only information projected onto $\bupSigma$.
Passing to $\bigupsigma$ involves forgetting that $\FrM$ was the source of this information, now to be regarded as set up from intrinsic first principles.  
{\it Forget} involves forgetting that $\langle \FrM, \bupSigma \rangle$ contains a particular picked-out hypersurface as a slice.\footnote{A {\it forgetful map} 
is one which `forgets' (or `strips off') some of the layers of structure: $\phi: \langle \FrS, \sigma\rangle \rightarrow \FrS$.
There is corresponding loss of structure preservation in the maps associated with the latter equipped space.
In fact, this forgetful notion readily extends to a functorial notion in Category Theory.}  
%
Since $\FrM = \bupSigma \times \FrI$, $\bupSigma$ and $\bigupsigma$ are for now merely related by the identity map.  
On the other hand, {\it Embed} involves allowing for $\bupSigma$ to be treated as a hypersurface within some surrounding $\FrM$; this map is denoted by $\Phi$.  
Since $\Phi: \bigupsigma \rightarrow \FrM$, the embeddings in question are homeomorphisms.
The 1-to-1 ness within that statement guarantees that the spatial hypersurface $\bupSigma$ does not intersect itself, unlike Fig \ref{Top-Emb-Fol-Push-Pull}.c).
See Fig \ref{Top-Emb-Fol-Push-Pull}.a) for how the above maps fit together.  

{            \begin{figure}[ht]
\centering
\includegraphics[width=0.8\textwidth]{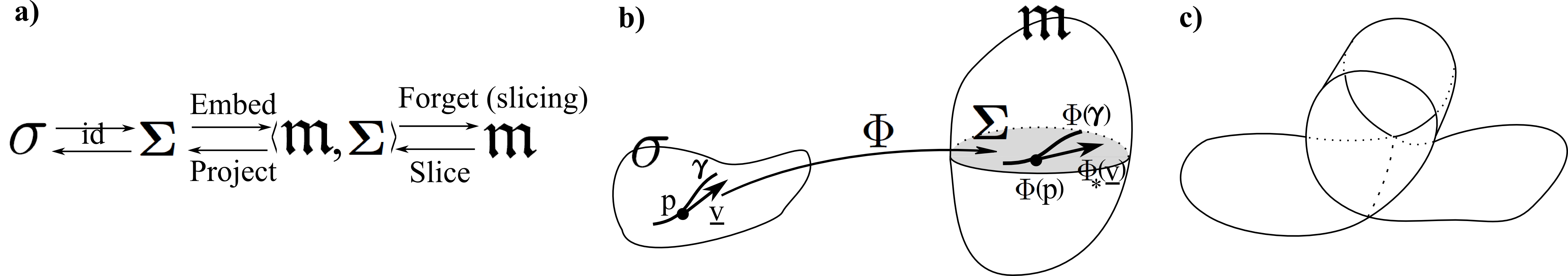}
\caption[Text der im Bilderverzeichnis auftaucht]{        \footnotesize{a) The detailed sense in which (suitable extensions of) `embed' and `slice and project' 
form a 2-way route at the level of topological manifolds.  
b) Given an identity-and-embedding $\Phi \, \circ $ id (which we will take $\Phi$ to suffice to denote), 
and allowing for the spaces in question to possess differentiable structure, the following additional maps are established.  
The corresponding \cite{I93, Gour} push-forward $\Phi_*\underline{v}$ of a tangent vector $\underline{v} \in \FrT_p(\bigupsigma)$       to         a curve $\gamma$       in $\bigupsigma$ 
                                                                 is then a tangent vector               $\in \FrT_{\Phi(p)}(\bupSigma)$ to the image curve $\Phi(\gamma)$ in $\bupSigma$. 
In the opposite direction -- projection-and-forgetting -- $\Phi$ induces a pull-back 
$\Phi^*: \FrT^*_{\Phi(p)}(\bupSigma) \rightarrow \FrT^*_p(\bigupsigma)$ between the space of 1-form linear maps. 
%
%
%
c) Embeddings preclude self-intersections.    
}  } 
\label{Top-Emb-Fol-Push-Pull}\end{figure}          }

\ni Allowing for differentiable manifold structure as well, Fig \ref{Top-Emb-Fol-Push-Pull}.b) associates embedding and slicing with the differential-geometric notions 
of push-forward and pull-back \cite{Stewart}.

\subsection{ii. Metric level}\label{Hyp-Geom-2}

The situation is more complicated at the metric level. 
One fundamental reason for these complications is that {\sl each} of $\FrM$, $\bupSigma$ and $\bigupsigma$ carries its own metric, moreover with the first two bearing relation.  
Let us begin considering this by splitting the metric geometry $\langle \FrM, \bg \rangle$ with respect to $\bupSigma$ into of tangential and normal parts:

\ni\beq
\FrT_{\sp}(\FrM) = \FrT_{\sp}(\bupSigma) \oplus \FrN \mbox{ } , 
\eeq
for $\FrN$ the space spanned by the normal vector field $\mn^{\sfA}$.
This is normal to the corresponding {\it isometric embedding}: $\Phi: \bupSigma \rightarrow \FrM$. 
[This is $\Phi_{\st}$ for some particular fixed value of $\mt$.]
Using $\Phi^{\mu} := \vec{X}^{\mu}(\Phi(\ux))$ for $\vec{X}^{\mu}$ a coordinate system on $\FrM$ \cite{Kuchar76I, IK85, I93, Giu15}, this is defined by 

\ni\be
\mn_{\mu}(\ux; \Phi] \, \Phi^{\mu}\mbox{}_{,i}(\ux) = 0  \mbox{ } , 
\label{n-1}
\ee

\ni\be
\mg^{\mu\nu}({\Phi}(\ux)) \, \mn_{\mu}(x^{\gamma}, \Phi] \, \mn_{\nu}(\ux;\Phi] = - 1 \mbox{ } \mbox{  } \forall \, x \in \bupSigma \mbox{ } .
\label{n-2}
\ee
Here, (\ref{n-1}) is normality in the sense of `being perpendicular' to the hypersurface.
The {\it induced metric} can then be interpreted as the pull-back $\bh := \Phi^*\bg$.
In components, using a hypersurface-adapted coordinate system \ux, $t$, with $t = t_1$ picking out the hypersurface itself, 

\ni \beq
\mh_{ij}(\ux, t_1) := \mh_{ij}(\ux, \Phi_{t_1}] = (\Phi^*\bg)_{ij}(\ux, t_1) = (\Phi^*_{t_1} \bg)_{ij}(\ux) = 
\mg_{\mu\nu}(\Phi(\ux, t_1)) \, \Phi^{\mu}\mbox{}_{,i}(\ux, t_1) \, \Phi^{\nu}\mbox{}_{,j} (\ux, t_1) \mbox{ } . 
\label{Phi-ab}  
\eeq
\mbox{ } \mbox{ } A second fundamental reason for the metric-level version's greater complication is that one now has not only 

\ni $\bh \in$ $\mbox{Riem}(\bupSigma)$ to contend with, but also a notion of extrinsic curvature $\bK$, whose possible values form the space Sym($\bupSigma)$.
The possible extrinsic curvatures form the space of symmetric 2-tensors on $\bupSigma$.
The complication is then that one needs to associate a $\bK$ to each $\bh$.\footnote{One formulation of this involves the tangent bundle space $\FrT$(Riem(\bupSigma))  
of extrinsic curvatures over the base space $\mbox{Riem}(\bupSigma)$. 
Another formulation involves using $\bp$ in place of $\bK$. 
In the latter case, the Sym involved is a space of symmetric 2-tensor {\sl densities} and the corresponding cotangent bundle space is $\FrT^*(\mbox{Riem}(\bupSigma))$.\label{Bun-Foo}}

Then {\it Slice} involves identifying a particular slice $\langle \bupSigma, \bh, \bK \rangle$ in $\langle \FrM, \bg \rangle$.   
{\it Project} involves keeping only information projected onto $\langle \bupSigma, \bh, \bK \rangle$.
Passing to $\bigupsigma$ involves forgetting that $\FrM$ was the source of this information, now including forgetting $\bK$ since that is not intrinsic to $\bigupsigma$.  
{\it Forget} now involves forgetting that $\langle    \langle \FrM, \bM \rangle,    \langle \bupSigma, \bm, \bK \rangle    \rangle$ 
contains a particular picked-out hypersurface                                         $\langle \bupSigma, \bm, \bK\rangle$             as a slice.
Other reasons for the projection step being less trivial than at the topological level include needing firstly to derive the Gauss--Codazzi relations 
(\ref{Gauss-as-proj}--\ref{Cod-as-proj}) by projections.
Secondly, one then needs to solve these equations in order to obtain suitable $(\bh, \bK)$ pairs.
Moreover, $\langle \bupSigma, \bh \rangle$ and $\langle \bigupsigma, \bh \rangle$ are {\sl not} now related by the identity map.
For not all $\bh$ one can place on $\bigupsigma$ are necessarily isometrically embeddable into $\langle \FrM, \bg \rangle$.  
Hence a nontrivial inclusion map is required.\footnote{For $\FrW \subset \FrU$, 
the corresponding {\it inclusion map} is the injection $j: \FrW \rightarrow \FrU$ with $j(w) = w \mbox{ } \mbox{  } \forall \, w \in \FrW$.}  
%
{\it Embed} now involves allowing for $\langle \bupSigma, \bm \rangle$ to be treated as a hypersurface within some surrounding $\langle \FrM, \bM \rangle$; 
this map continues to be denoted by $\Phi$.

\subsection{Foliation in terms of a decorated chart}

%
Let us first generalize Sec \ref{Gdyn}'s definition of {\it foliation} to $\FrF = \{ \FrL_{\sfA}\}_{\sfA \in A}$. 
In this article, this denotes a decomposition of an $4-d$ manifold ${\FrM}$ into a disjoint union of connected $3-d$ subsets 
(the {\it leaves} $\FrL_{\sfA}$ of the codimension $1 = 4 - 3$ foliation) such that the following holds. 
$m \in \FrM$ has a neighbourhood $\FrU$ with coordinates ($x^0, ..., x^3$). 
I.e. $\FrU \rightarrow \mathbb{R}^4$ such that for each leaf $\FrL_{\sfA}$ the components of $\FrU \bigcap \FrL_{\sfA}$ are described by $x^0$ constant:  
the obvious extension of Fig \ref{Infinitesimal-Fol}.b).

\section{ADM versus A kinematics for foliations}\label{Fol-ADM-A}

%
The foliation $\Phi_{t}$ is a map (in fact a diffeomorphism) $\bupSigma \times \mathbb{R} \rightarrow \FrM$.  
%
%
Thus its inverse $\Phi^{-1}\mbox{}_{t} : \FrM \rightarrow \bupSigma \times \mathbb{R}$ is also a diffeomorphism. 
The form of the latter is then \cite{I93} $\Phi_{t}\mbox{}^{-1}(\vec{X}) = (\upsigma(\vec{X}),\uptau(\vec{X})) \rightarrow \bupSigma \times \mathbb{R}$.   
$\uptau : \FrM \rightarrow \mathbb{R} $ is here a global time function, 
providing the time $t := \{\uptau(\Phi_{t}(x)) = t \mbox{ } \mbox{  } \forall \,  x \in \bupSigma\}$  in the sense of natural time parameter corresponding to the foliation. 
Isham \cite{I93} cautions, however of the artificiality of such a `definition of time' from an operational perspective, i.e. concerning its measurability and associated clock manufacture.  
On the other hand, $\upsigma : \FrM \rightarrow \bupSigma$ is a `space map'.\footnote{See e.g. \cite{KK02-Kouletsis08} for further developments of this in a histories theory context.}

\mbox{ } 

\ni For each $\ux \in \bupSigma$, the map $\Phi_{\ux} : \mathbb{R} \rightarrow \FrM$ defined by $t \mapsto \Phi(\ux, t) := \Phi_{t}(\ux)$ is a curve in $\FrM$. 
This has a corresponding to it a 1-parameter family of tangent vectors on $\FrM$.

\mbox{ }

\ni In the ADM version, let us denote these by 

\ni\beq 
\dot{\Phi}_{\ux}(t)  \mbox{ whose components are } \mbox{ } \dot{\Phi}^{\mu}_{\ux}(t) =  \dot{\Phi}^{\mu}(\ux, t)/\pa t \mbox{ } . 
\label{dot-Phi}
\eeq
The corresponding vector field is the {\it deformation vector field}; Fig \ref{I-1-2} gives the corresponding flow lines.  
This has the following duality (the current Article's fourth distinct type of duality) 
for each $\ux \in \bigupsigma$, $\dot{\Phi}_{\ux}(t)$ is a vector in $\FrT_{\Phi(\ux,t)}(\FrM)$ at the point $\Phi(\ux,t)$ in $\FrM$. 
Isham counsels \cite{I85, I93} that this object is best regarded as an element of $\FrT_{\Phi_t}(\mbox{Emb}(\bigupsigma, \FrM))$: the space of vectors tangent to 
the infinite-$d$ manifold $\mbox{Emb}(\bigupsigma, \FrM)$ of embeddings of $\bigupsigma$ {\it in} $\FrM$ {at the particular embedding $\Phi_t$.   

Moreover, the deformation vector field is a reinterpretation of Sec \ref{Gdyn}'s time flow vector field $\mt^{\mu}$ according to

\ni\beq
\mt^{\mu}(\vec{X}) = \left. \dot{\Phi}^{\mu}(\ux, t) \right|_{\vec{X} = \vec{X}(\ux, t)} \mbox{ } , 
\eeq
corresponding to viewing it acting on a slice or leaf. 
%

\mbox{ }

\ni \underline{Highlight 14}. In the A version, let us encode the above structure instead as the TRi-covector 

\ni\beq
\pa \Phi_{\ux}(t)  \mbox{ whose components are } \mbox{ } \pa \Phi^{\mu}_{\ux}(t) = \pa\Phi^{\mu}(\ux, t) \mbox{ } . 
\label{Tang}
\eeq
The corresponding TRi-covector vector field is then the {\it TRi-covector version of} the deformation vector field; Fig \ref{I-1-2}.b) depicts the corresponding flow lines.
Note that $\pa \ms^{\mu}$ is more primary in this approach than $\pa \mF^i$ and $\pa \mI$ (or $\pa \mX^i$ and $\pa \mT$). 
Thus, it is the first object to be allocated nontrivial TRi homothetic weight and to set the `vector or covector' sign convention and the size convention for the unit weight \cite{TRiPoD}.
Also the above-mentioned duality now takes the following form. 
For each $\ux \in \bigupsigma$, $\pa{\Phi}_{\ux}(t)$ is a vector in $\FrT_{\Phi(\ux, t)}(\FrM)$ at the point $\Phi(\ux, t)$ in $\FrM$. 
Moreover, the TRi-covector deformation vector field is a reinterpretation of the TRi-covector time flow vector field $\pa \ms^{\mu}$ according to

\ni \be
\pa\ms^{\mu}(X) = \left. \pa \mX^{\mu}(\ux, t) \right|_{\vec{\sX} = \vec{\sX}(\ux, t)}  \mbox{ } .
\ee
again corresponding to viewing this as acting on a slice or leaf.						  
Thus in approaches involving deformation primality, the previous comment about the time flow being the first nontrivial TRi homothetic weight quantity to be encountered 
carries over to the elsewise also primary TRi-covector deformation vector.

{            \begin{figure}[ht]
\centering
\includegraphics[width=1.0\textwidth]{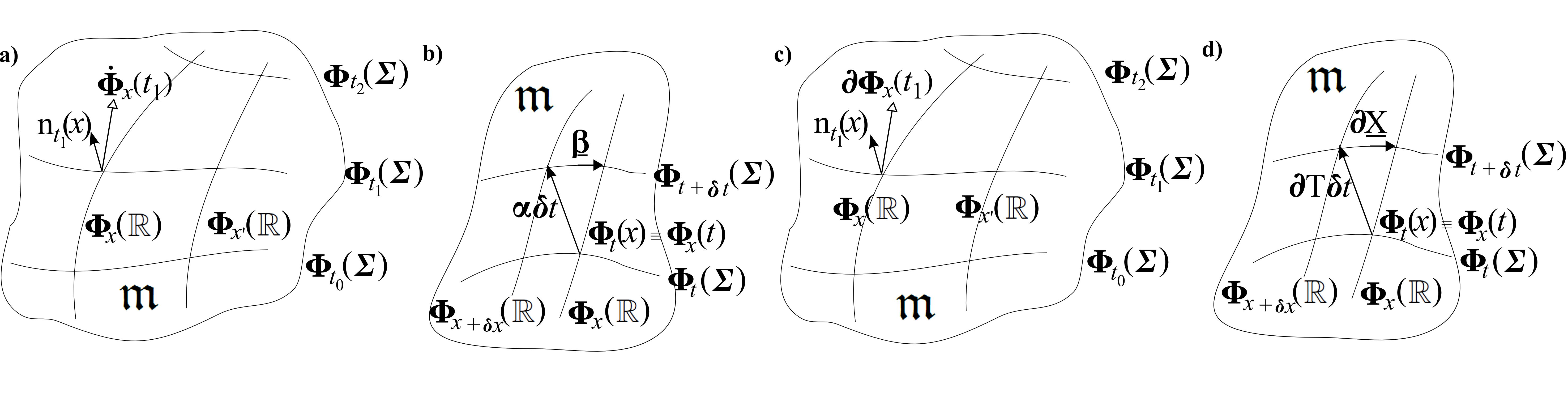}
\caption[Text der im Bilderverzeichnis auftaucht]{        \footnotesize{a) The flow lines of the foliation of $\FrM$ \cite{I93}.
Also, $\mn_{t_1}$ is the normal vector field on the hypersurface $\Phi_{t_1}(\bupSigma) \rightarrow \FrM$ and extending along the flowlines, 
one has the normal vector field $\mn_t$ to the whole foliation.
b) The lapse and shift in the context of a foliation \cite{IK85, I93}.
c) The flow lines of the TRi foliation of $\FrM$.
d) The differentials of the instant and of the frame in the context of a TRi foliation.}  } 
\label{I-1-2}\end{figure}          }

\ni The functional derivative of $\mh_{ab}(\ux; \Phi]$  with respect to $\Phi$ projected along $\mn^{\mu}$ is of value in considering dynamical evolution.
Computationally, \cite{Kuchar74, HKT} this takes the form

\ni\beq
\mn^{\mu}(\ux; \Phi] \, \frac{\delta}{\delta \Phi^{\mu}(\ux))\mh_{ab}(\ux^{\prime}, \Phi]} =
                                        2 \, \mK_{ab}(\ux, \Phi] \, \updelta(\ux - \ux^{\prime}) \mbox{ } ,
\eeq
for $\bK$ the extrinsic curvature of the hypersurface $\Phi(\bupSigma)$, which in this formulation is given by [c.f. (\ref{Phi-ab})]

\ni\beq
\mK_{ab}(\ux; \Phi] := - \nabla_{\mu}\mn_{\nu}(\ux; \Phi] \, \Phi^{\mu}\mbox{}_{,a} \, (\ux)\Phi^{\nu}\mbox{}_{,b}(\ux) \mbox{ } .
\label{K-Emb}
\eeq 
Also, $\nabla_{\mu}\mn_{\nu}(\ux; \Phi]$ is here the covariant derivative obtained by parallel transporting the cotangent vector 

\ni $\mn(\ux, \Phi] \in \FrT^*_{\Phi(\ux)}(\ux)\FrM$ along the hypersurface $\Phi(\bupSigma)$ using $\FrM$'s metric $\bg$.  
The deformation vector can then be decomposed into one piece that lies along the hypersurface $\Phi_{\st_1}(\bupSigma)$ and another parallel to $\mn_{\st}$. 

\mbox{ }

\ni Then in the ADM formulation, (\ref{dot-Phi}) is expanded out as

\ni\beq
\dot{\Phi}^{\mu}(\ux, t_1) = \upalpha(\ux, t_1) \,  \mg^{\mu\nu}(\Phi(\ux, t_1)) \, \mn_{\nu}(\ux, t_1) + 
                                       \upbeta^a(\ux, t_1) \, \Phi^{\mu}\mbox{}_{,a}(\ux, t_1) \mbox{ } , 
\eeq
using $\mn^{\mu}(\ux, t)$ as shorthand for $\mn^{\mu}(\ux, \Phi_{t}]$. 
From a more minimalist perspective, note that lapse and shift remain meaningful for just a pair of neighbouring slices.
(This is now indexed by $t_1$: the value on the initial slice, and interpreted in terms of a single embedding $\Phi_{t_1}$ corresponding to this slice.)  
Indeed, one can go as far as reinterpreting the shift in terms of coordinate changes on a single hypersurface.\footnote{Note that this is Lie dragging within a single slice, 
as opposed to a point identification map between two neighbouring slices. \label{2-barrel}}

\mbox{ }

\ni \underline{Highlight 15}. On the other hand, in the A formulation, the Tri covector (\ref{Tang}) is expanded out as 

\ni\beq
\pa{\Phi}^{\mu}(\ux, t_1) = \pa \mT(\ux, t_1)   \, \mg^{\mu\nu}(\Phi(\ux, t_1)) \, \mn_{\nu}(\ux, t_1) + 
                            \pa \mX^a(\ux, t_1) \, \Phi^{\mu}\mbox{}_{,a}(\ux, t_1) \mbox{ } .  
\eeq
From a more minimalist perspective, note that differential of the instant and differential of the frame remain meaningful for just a pair of neighbouring slices.
(This is again indexed by $t_1$.)  
Indeed, one can interpret the differential of the frame, in terms of coordinate changes on a single hypersurface.$^{\ref{2-barrel}}$

\mbox{ } 

\ni Then from the spacetime perspective $\upalpha$ and $\upbeta^i$ depend on the spacetime metric $\bg$ as well as on the foliation (a partly invertible relationship \cite{I93}).
For a fixed foliation, $\upalpha$ and $\upbeta^i$ are identified with pieces of $\bg$.
E.g. this can be formulated using the pull-back $\Phi^*(\bg)$ of $\bg$ 
by the foliation $\Phi: \bupSigma \times \mathbb{R} \rightarrow \FrM$ in coordinates $\vec{X}^{\mu}, \, \mu = 0 ... 3$, on $\bupSigma \times \mathbb{R}$ 
that is adapted to the product structure: $\vec{X}^{\mu = 0}(\ux, t) = t$.
Here \cite{I93} $\vec{X}^{\mu = 1, 2, 3}(\ux, t) = \ux^{\mu = 1, 2, 3}(\ux)$ for $x^a, \, a = 1, 2, 3$ some coordinate system on $\bupSigma$. 

\mbox{ } 

\ni Then in the ADM formulation, the components of $\Phi^*(\bg)$ are (\ref{Phi-ab}), 

\ni\beq
(\Phi^*\bg)_{0a}(\ux, \mt) = \upbeta^{b}(\ux, \mt) \, \mh_{ab}(\ux, t) \mbox{ } , 
\label{Phi-0a}
\eeq

\ni\beq
(\Phi^*\bg)_{00}(\ux, t) = \upbeta^{a}(\ux, t) \, \upbeta^b(\ux, t) \, \mg^{ab}(\ux, t) - \upalpha(\ux, t)^2 \mbox{ } .  
\label{Phi-00}
\eeq
\ni \underline{Highlight 16}. On the other hand, in the A formulation, the components of $\Phi^*(\bg)$ are (\ref{Phi-ab}), the TRi-covector

\ni\beq
(\Phi^*\bg)_{0a}(\ux, t) = \pa \mX^{b}(\ux, t) \, \mh_{ab}(\ux, t) \mbox{ } , 
\label{TRi-Phi-0a}
\eeq
and the TRi 2-tensor

\ni\beq
(\Phi^*\bg)_{00}(\ux, t) = \pa \mX^{a}(\ux, t) \, \pa \mX^b(\ux, t) \, \mg^{ab}(\ux, t) - \pa \mT(\ux, t)^2 \mbox{ } .  
\label{TRi-Phi-00}
\eeq
In this way, the A conception of the split can be replaced by one based on TRi-foliations. 
This parallels the replacement of the ADM split by one based on foliations in \cite{Kuchar76I, IK85, I93}.
This is $\Phi^{\sr\se\sf}: \bupSigma \times \mathbb{R} \rightarrow \FrM$ with respect to some choice of reference foliation.

\section{TRi Refoliation Invariance}\label{TRi-RI}

%
Having separated out 3-space, 1 hypersurface, and 2 or more hypersurface properties, now identify Refoliation Invariance as a comparison of triples of hypersurfaces.
This matches the mathematical notion of commutator (Fig \ref{DiracAlgebroid4}.a), 
which the algebraic commutator resolution of the matter by Teitelboim \cite{Tei73} indeed makes use of.
In particular, it is comparing two triples starting from the same object, each of which triples involves applying the same two operations but in opposite orders. 
The question then is whether the outcome of these two different orders is the same. 
The mathematically strongest notion of sameness here is identity, however the two outcomes being out by a physically irrelevant transformation is also here acceptable. 
See \cite{AM13} for the corresponding interpretation of the Dirac algebroid's other two Poisson brackets.

Additionally, in the current Sec we pass from the usual presentation of the Dirac algebroid (\ref{Mom,Mom}-\ref{Ham,Ham}) to the TRi-smeared version (\ref{TRi-Mom,Mom}-\ref{TRi-Ham,Ham}). 
Due to this, Fig \ref{DiracAlgebroid4}.b) replaces Fig \ref{Refol-4-SCPs}.b).

{            \begin{figure}[ht]
\centering
\includegraphics[width= 0.4 \textwidth]{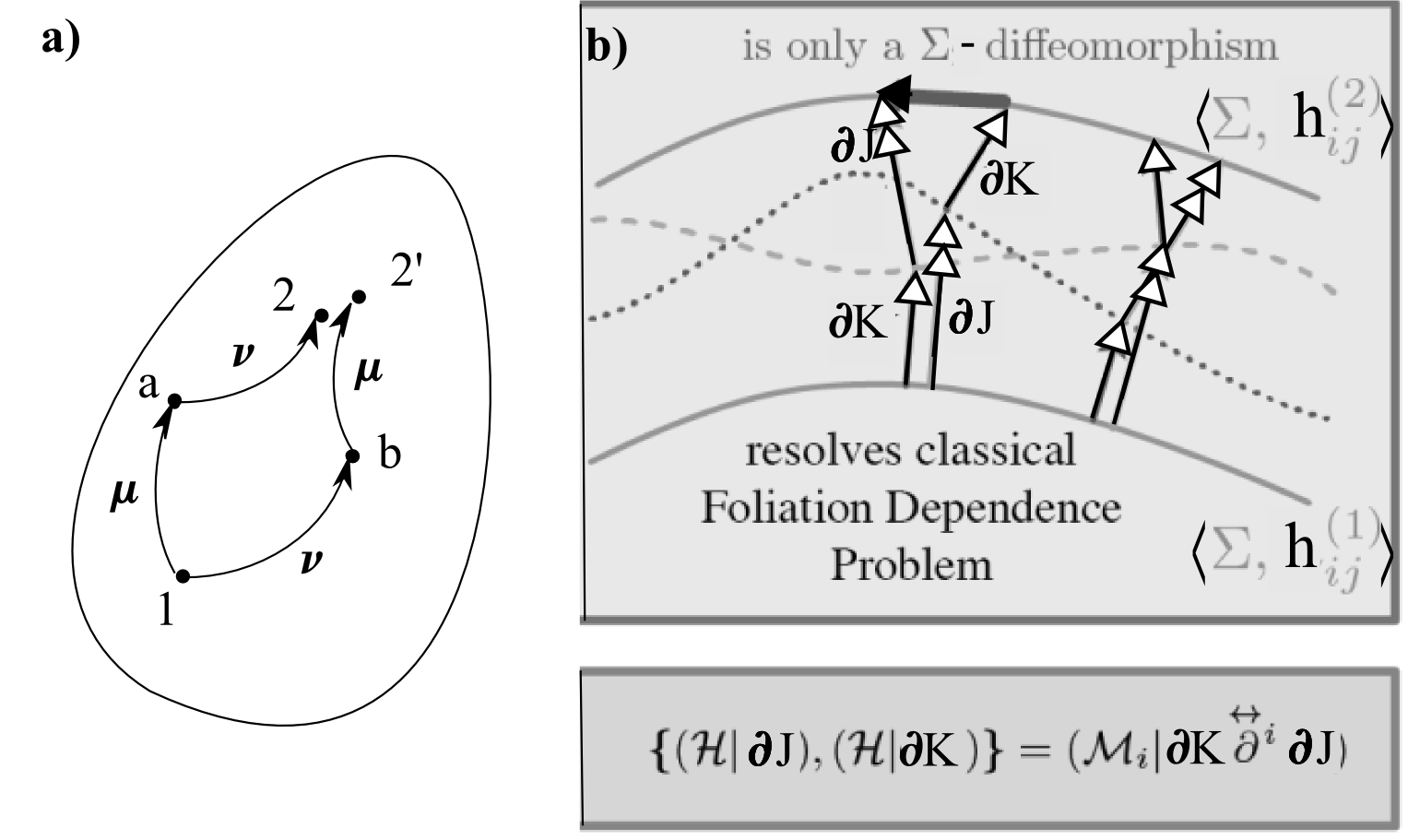} 
\caption[Text der im Bilderverzeichnis auftaucht]{        \footnotesize{a) The geometrical commutator as a comparison of two triples of objects, 
resulting from applying two transformations $\mu$, $\nu$ in either order to an initial object $1$. 
If $2 = 2^{\prime}$, $\mu$ and $\nu$ commute.  
b) In the case in hand, the points depicted in a) are in a suitable bundle$^{\ref{Bun-Foo}}$ over $\mbox{Riem}(\bupSigma)$, so as to represent the hypersurfaces indicated.  
The correspondence is $1$ to $\langle \bupSigma, \mh^{(1)}_{ij} \rangle$, $a$ to the dashed hypersurface, $b$ to the dotted hypersurface.
Then the algebraic commutator relation at the bottom of the Figure ensures the TRi version of Refoliation Invariance for classical GR in close parallel to Teitelboim's result \cite{Tei73}.

\mbox{ } 

\ni\underline{Highlight 17}. By this, $2 = \langle \bupSigma, \mh^{(1)}_{ij} \rangle  = 2^{\prime}$ modulo just a diffeomorphism of the final hypersurface, 
now associated with the TRi smearing $\pa\mK\stackrel{\leftrightarrow}{\pa^i}\pa\mJ$.               } }
\label{DiracAlgebroid4} \end{figure}          }

\ni Note that Refoliation Invariance is also known as {\it path independence} \cite{Bubble, HKT}.  
%
%
To be clear, this refers to a path in configuration space rather than in spacetime, 
and is in that context not only a `dimension--codimension' type of duality but also a `ray--wavefront' one as well.

\section{Duality between many-fingered time and bubble time}\label{Bub-Finger}

Multiple choices of timefunction are valid in GR.  
Each corresponds to a foliation.
The first of these reflects GR coordinate time's multiplicity and what becomes of this upon performing a space-time split, giving a `many-fingered' notion of time.  
Time is moreover local in GR: in some places at a given time, a choice of finger is longer than at other places (Fig \ref{Bubble-Finger}.b).
The instant of time is a slice, with a continuous sequence of non-intersecting slices forming a foliation as per Sec \ref{Fol-2}.
Then thinking of a particular slice as the surface of a bubble, one encounters the further notion of a bubble deforming\footnote{This notion of deformation 
is the same as that in Secs \ref{Fol-ADM-A} and \ref{HKT}.}  
under evolution due to the above local aspect of GR time.  
The `many-fingered time' to `many bubble deformations time' is moreover a `ray--wavefront' type of duality in spacetime (Fig \ref{Bubble-Finger}.a). 
%
{            \begin{figure}[ht]
\centering
\includegraphics[width=0.45\textwidth]{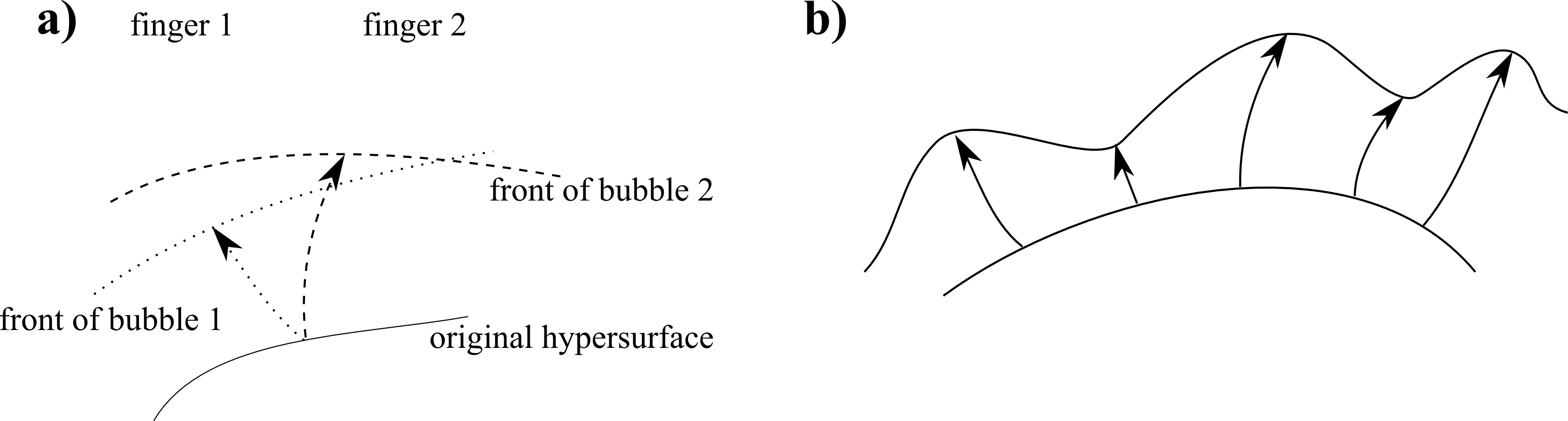}
\caption[Text der im Bilderverzeichnis auftaucht]{        \footnotesize{a) Each finger of time corresponds to deforming the original hypersurface to a different bubble front.
b) Note that each of these is additionally a {\sl field theoretic} notion, meaning that the lengths of a given choice of finger in general vary from point to point.
Thus it corresponds to an outbreak of small bubble deformations. }  } 
\label{Bubble-Finger}\end{figure}          }

N.B. that bubble time is more generally a {\it field-theoretic} feature rather than just a geometrodynamical one.
In the field-theoretic context, it was developed by Weiss \cite{Weiss} and by Tomonaga \cite{Tomonaga}; 
as a functional integral method, this eventually bears close relation to path integral formulations.
Indeed, the bubble time notion entered canonical GR through Dirac being aware \cite{Dirac48b} of the preceding development in field theory.
The bubble time notion's relation to deformations in geometrodynamics was further developed in \cite{Bubble, Kuchar73, HKT}.  

\mbox{ } 

\ni That a bubble time presentation covers an infinity of ADM presentations at once.
This refers to ADM involving particular local choice of coordinates, by which Refoliation Invariance ceases to be manifest. 
Dirac's own approach avoided this, while not being the same as Kucha\v{r}'s bubble time approach either; the two are canonically related \cite{Bubble}.  
Use of the bubble time notion in geometrodynamics is a `covariantizing' feature -- a mathematical implementation \cite{Bubble} of a prior insight of Wheeler's.   
This is attained through the formalism's many fingered dual aspect considering all coordinate times at once.
Alternatively, this formalism's bubble aspect considers all foliations at once. 
Consequently, it also considers whichever families of arbitrarily moving observers at once (by Sec \ref{Obs}).    

\mbox{ }

\ni \underline{Highlight 18}. The above coordinate fixing aspect of ADM carries over to the case in which the ADM split is substituted for the A split.

\ni \underline{Highlight 19} However, the above bubble time reformulation which \K applied to free ADM's original approach from this coordinate fixing aspect 
can also be applied to free the A parallel \cite{FEPI} of ADM's original approach from its own coordinate fixing aspect.
In particular, \cite{Bubble} eliminates lapse and shift, corresponding to another formulation of the thin sandwich.
The corresponding TRi version then eliminates cyclic differential of the instant and of the frame, corresponding to another formulation of the TRi Machian Thin Sandwich.

\section{GR: from deformation first principles}\label{HKT} 

In addressing to Wheeler's question (\ref{Wheeler-Q}), Hojman, Kucha\v{r} and Teitelboim (HKT) presupposed embeddability into spacetime out of following Wheeler's further advice 
\cite{Battelle}: {\it``the central starting point in the proposed derivation would necessarily seem to be imbeddability"} (sic). 
\mbox{ } 

\ni They start from assigning primality to hypersurface deformations.
This rests upon the existence of 3-$d$ spacelike hypersurfaces described by Riemannian geometry and the embeddability of these into (conventional 4-$d$, semi-Riemannian) spacetime.
The actions of the generators of the pure deformations and the stretches are as per Fig \ref{def}. 

{\begin{figure}[ht]
\centering
\includegraphics[width=0.6\textwidth]{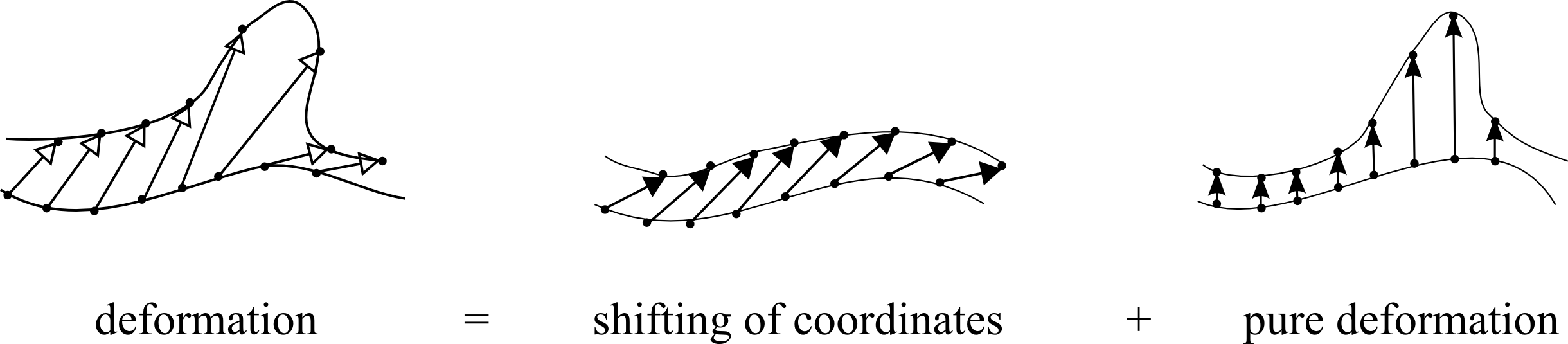}
\caption[Text der im Bilderverzeichnis auftaucht]{\footnotesize{The general deformation of a surface decomposes into a pure stretch of a surface and a pure deformation of that surface. 
These are generated by $\scP\scU\scR\scE$ and $\scS\scH\scU\scF\scF\scL\scE_i$ respectively.
%
%
        The action of $\scS\scH\scU\scF\scF\scL\scE_i$ is a mere shuffling around of points within a given hypersurface, 
whereas the action of $\scP\scU\scR\scE$               involves deforming that hypersurface itself.}} 
\label{def}\end{figure}} 

\ni Next, evaluating the Poisson brackets of these generators gives the {\it deformation algebroid} \cite{Tei73b} 

\ni\be
\mbox{\bf \{} \scS\scH\scU\scF\scF\scL\scE_i(\ux) \mbox{\bf ,} \scS\scH\scU\scF\scF\scL\scE_j(\ux^{\prime}) \mbox{\bf \}} = 
\scS\scH\scU\scF\scF\scL\scE_i(\ux^{\prime})\updelta_{,j}(\ux, \ux^{\prime}) + 
\scS\scH\scU\scF\scF\scL\scE_j(\ux)\updelta_{,i}(\ux ,\ux^{\prime} )            \mbox{ } ,
\label{DeformAlgebra1}
\ee

\ni\be
\mbox{\bf \{ }\scS\scH\scU\scF\scF\scL\scE_i(\ux)\mbox{\bf ,} \scP\scU\scR\scE(\ux^{\prime}) \mbox{\bf \}} = 
\scP\scU\scR\scE(\ux)\updelta_{,i}(\ux,\ux^{\prime}) \mbox{ } , \mbox{ } 
\label{DeformAlgebra2}
\ee

\ni\be
\mbox{\bf \{ } \scP\scU\scR\scE(\ux) \mbox{\bf , } \scP\scU\scR\scE(\ux^{\prime}) \mbox{\bf \}} = 
\mh^{ab}(\ux)          \scS\scH\scU\scF\scF\scL\scE_j(\ux) \updelta_{,i}(\ux, \ux^{\prime}) + 
\mh^{ab}(\ux^{\prime}) \scS\scH\scU\scF\scF\scL\scE_j(\ux^{\prime}) \updelta_{,i}(\ux, \ux^{\prime}) \mbox{ } .
\label{DeformAlgebra3}
\ee
I give this as unsmeared, though it was implicitly derived with the plain smearings corresponding to the ADM formulation. 

\mbox{ }

\ni \underline{Highlight 20} The above can however be straightforwardly rederived using the A formulation's TRi smearings. 

\mbox{ }

\ni To implement the deformation algebra as a first principle for prospective gravitational theories, HKT evoke the {\it representation postulate}.
I.e. that for conventional spacetime to be produced,                         $\scH^{\st\sr\si\sa\sll}$ and $\scM_{i}^{\st\sr\si\sa\sll}$ constraints for these theories are to take a form 
such that they close in the same form as the deformation algebroid formed by $\scP\scU\scR\scE$        and $\scS\scH\scU\scF\scF\scL\scE_i$. 
GR indeed satisfies this because the Dirac algebroid is of this form; the point however is inserting much more general ans\"{a}tze and proving that these narrow down to the GR case alone.

A subsidiary assumption evoked by HKT is {\it locality}: that the metric is to be only locally affected by a pure deformation.
For the usual reasons in brackets structure considerations, $\scM_i^{\st\sr\si\sa\sll}$ is fixed to be the GR $\scM_i$ by the first bracket.
The second bracket solely restricts $\scH^{\st\sr\si\sa\sll}$ to be a scalar density of weight 1, whereas the lion's share of the calculation involves the last bracket.  
Here, HKT's result involves assuming locality and 2 degrees of freedom per space point. 
It proceeds by induction and by evoking the 3-$d$ counterpart of Lovelock's Theorem.
This results in the GR form of $\scH$ including cosmological constant term, 
alongside noting that altering the signature also preserves the result up to at most a sign in the algebraic structure of the deformation generators' Poisson brackets.  

\mbox{ } 

\ni \underline{Highlight 21} These results of HKT can also be straightforwardly recast in A split terms.  
However, this is not used in the Relational Approach since that {\sl still} presupposes spacetime, 
whereas the Relational Approach has its own procedure as per Sec \ref{TRi-SCP} which is additionally a bona fide Spacetime Construction.

\mbox{ } 

\ni Teitelboim \cite{Teitelboim} also succeeded in including the habitually used fundamental bosonic matter $\uppsi$ into HKT's scheme.
Here the Einstein--matter system's Hamiltonian and momentum constraints are of the form

\ni
\be
\scH_{\sg\uppsi} = \scH_{\sg} + \scH_{\uppsi} \mbox{ } \mbox{ and } \mbox{ } ( \scM_{\sg\uppsi})_i = ( \scM_{\sg})_i + (\scM_{\uppsi})_i \mbox{ } .
\ee
The representation postulate idea then extends additively to the constraints' {\sl matter contributions}.
I.e. these {\sl separately} obey the Dirac algebroid.

\mbox{ }  

\ni \underline{Highlight 22} These matter results of Teitelboim can also be derived in TRi terms based upon the A split.  

\mbox{ } 

\ni HKT's work has been further compared with the Relational Approach in \cite{Phan, AM13}.
It has also been enlighteningly discussed in e.g. \cite{Kouletsis, KieferBook, Giu09}.
The first of these is in a histories-theoretic context while the last of these touches upon the extent to which Ashtekar variables approaches have a counterpart.

\section{Universal hypersurface kinematics}\label{Hyp-Kin}

The split with respect to a hypersurface $\bupSigma$ of the spacetime covariant derivative $\nabla_{\mu}$ acting on a general spacetime tensor field does not just produce, 
the obvious spatial covariant derivative $\mD_i$ \cite{Kuchar76II, Giu15}.  
It additionally produces pieces with the following three universal hypersurface-geometrical interpretations \cite{Kuchar76I, Kuchar76II, Kuchar76III, Kuchar77}.\footnote{Intuitively, 
these relations come about because spacetime derivatives are not equal to spatial derivatives. 
This is because the former have extra connection components, which the current formulation then interprets geometrically from the perspective of the hypersurface.}

\subsection{ADM split version}\label{ADM-Hyp-Kin}

\ni 1) Hypersurface derivatives $\Circ_{\vec{\upbeta}}$, as already encountered in Sec \ref{Gdyn}, implementing `{\it shift kinematcs}'.

\ni 2) {\it Tilts}, which are one part of the tilt--translation split, the {\it translation} part being such that $\upalpha(\mp) \neq 0$, $\{\pa_{i}\upalpha\}(\mp) = 0$, 
                                                                    and the      {\it tilt}        part       such that $\upalpha(\mp) = 0$, $\{\pa_{i}\upalpha\}(\mp) \neq 0$.
The suitability of this name is most simply envisaged in the context of the standard formulation of SR by a family of boosted 
observers on a flat spatial surface tilted at a fixed angle to the undeformed flat spatial surface.  
C.f. also the tilted flow in Fig \ref{Observer-Congruences}.b), in which manner tilt plays a role in theoretical cosmology. 

\ni 3) {\it Derivative couplings} are terms linear in each of the extrinsic curvature and the tensor field itself.  
Absence of such terms is known as the {\it Geometrodynamical Equivalence Principle} \cite{HKT}, which is a statement concerning minimal coupling (the complement of derivative coupling).

\subsection{1-form example}\label{1-form}

\ni For a 1-form $\mA^{\mu}$ (so this example is illustrating adjunction of matter) the 4-$d$ covariant derivative's pieces decompose as follows.

\ni\be
\nabla_{a}\mA_{\perp} = \mD_{a}\mA_{\perp} - \mK_{ab}\mA^{b} \mbox{ } ,
\label{ADMderivproj1}
\ee

\ni\be
\upalpha \nabla_{\perp} \mA_{a} = - \Circ_{\vec{\upbeta}}\mA_{a} - \upalpha \mK_{ab}\mA^{b} -  \mA_{\perp}\pa_{a}\upalpha \mbox{ } ,
\label{ADMderivproj2}
\ee

\ni\be
\nabla_{b} \mA_{a} = \mD_{b}\mA_{a} - \mA_{\perp}\mK_{ab} \mbox{ } ,
\label{ADMderivproj3}
\ee

\ni \be
\upalpha\nabla_{\perp}\mA_{\perp} = - \Circ_{\vec{\upbeta}}\mA_{\perp} - \mA^{a}\pa_{a}\upalpha  \mbox{ } , 
\label{ADMderivproj4}
\ee
Then on the right hand sides, (\ref{ADMderivproj2}) and (\ref{ADMderivproj4})'s first terms are hypersurface derivatives, their last terms are tilts, and
(\ref{ADMderivproj1}) and (\ref{ADMderivproj3})'s last terms and (\ref{ADMderivproj2})'s second term are derivative couplings.

\subsection{A split version: Machian hypersurface kinematics}\label{TRi-Hyp-Kin}

There are still three types, but their interpretations need to be redone \cite{AM13} from the perspective of spatial primality and this leads to two of them being renamed as well.  

\mbox{ } 

\ni 1) [\underline{Highlight 23}] {\it Best Matching derivatives} $\pa_{\underline{\sF}}$ (Sec \ref{TRi-CR}) replace the hypersurface derivatives $\Circ_{\vec{\upbeta}}$ (Sec \ref{CR}) 
they are dual to as per Sec \ref{A}.
These now implement `{\it spatial frame kinematics}'.  

\ni 2) [\underline{Highlight 24}] {\it Spatial gradients of the change of the instant} $\pa_i \pa \mI$ replace tilts $\pa_i \upalpha$.
The renaming involved here is part because because the name `tilt' is itself inspired by spacetime geometry; thus I reserve this terminology for the spacetime setting.
This is one part of the spatial gradient of the change of the instant--translation split. 
The TRi version of the translation is the part being such that $\pa\mI(\mp) \neq 0$, $\{\pa_{i}\pa\mI\}(\mp) = 0$, 
to the spatial gradient of the change of the instant part being such that $\pa\mI(\mp) = 0$, $\{\pa_{i}\pa\mI\}(\mp) \neq 0$.
Then after that point, the `proper time labels instants' duality converts this to the usual notion of tilt as recognizable from the above simple SR realization.
This is through the link of the observers' clocks' proper time exhibiting a constant gradient over space on the flat hypersurface tilted at a fixed angle.  

\ni 3) [\underline{Highlight 25}] {\it Derivative couplings} are as before, except that now the underlying formula for $\mK_{ij}$ is (\ref{K-Comp-A}) rather than (\ref{K-Comp-ADM}).
I.e. in spatially relational primary terms this is now interpreted as {\it comparison of each geometrodynamical change with the STLRC}.  

\mbox{ } 

\ni The above apply to reformulating the range of matter theories considered in \cite{Kuchar76III, Kuchar77, IN77}.
Also \cite{Van, Lan, Lan2} made use Kucha\v{r}'s hypersurface kinematics based on the ADM split followed by the analogue of BSW multiplier elimination of $\upalpha$ 
as a guide to which theories could be formulated from relational first principles. 
Then using the above TRi hypersurface kinematics based on the A split followed by cyclic differential Routhian reduction of $\mT$ recasts these calculations in a more relational form.

\subsection{Thin Sandwich Completion in terms of Machian kinematics}\label{TS-Complete} 

Here one finds that consistently accommodating a 3-tensor field sometimes requires associating it with further tensor fields in combinations such as 

\ni\beq
- \pa_{\underline{\sF}}\mA_{a} - \pa\mI \,\mK_{ab} \mA^{b} -  \mC \, \pa_{a}\pa\mI
\eeq
which, subsequently at the constructed spacetime level is interpreted as $\pa\mI \, \nabla_{\perp} \mA_{a}$ 
with $\mC$ recast in the role of the spacetime formulation's further piece $\mA_{\perp}$.

The idea here is to point out that Machian Thin Sandwich 3) contains construction of the Best-Matched derivative, whereas Machian Thin Sandwich 6) constructs $\mK_{ij}$.  
As such, postulating the following move completes the production of hypersurface kinematics.

\mbox{ } 

\ni Machian Thin Sandwich 7). Machian Thin Sandwich 4) further permits one to construct an emergent version of the spatial gradient of the change of instant, $\pa_b \, \pa\mI$.  

\mbox{ } 

\ni Thereby one can once again construct a wide range of spacetime geometrical objects as per above.

\mbox{ }

\ni For instance, re-running the above example gives the following two pairs, each consisting of one TRi-scalar and one TRi-covector

\ni\be
\nabla_{a}\mA_{\perp} = \mD_{a}\mA_{\perp} - \mK_{ab}\mA^{b} \mbox{ } ,
\label{Aderivproj1}
\ee

\ni\be
\pa \mI \, \nabla_{\perp} \mA_{a} = - \pa_{\underline{\sF}}\mA_{a} - \pa \mI \, \mK_{ab}\mA^{b} -  \mA_{\perp}\pa_{a}\pa \mI \mbox{ } ,
\label{Aderivproj2}
\ee

\ni\be\nabla_{b} \mA_{a} = \mD_{b}\mA_{a} - \mA_{\perp}\mK_{ab} \mbox{ } ,
\label{Aderivproj3}
\ee

\ni \be
\pa \mI \, \nabla_{\perp}\mA_{\perp} = - \pa_{\underline{\sF}}\mA_{\perp} - \mA^{a}\pa_{a}\pa \mI  \mbox{ } , 
\label{Aderivproj4}
\ee
Then on the right hand sides, (\ref{Aderivproj2}) and (\ref{Aderivproj4})'s first terms are Best Matching derivatives, 
their last terms are emergent spatial gradients of the instant, and(\ref{Aderivproj1}) and (\ref{Aderivproj3})'s last terms and (\ref{Aderivproj2})'s second term are derivative couplings.  
 
\mbox{ } 

\ni Parts 1)-4) and 6)-7) of the TRi Machian Thin Sandwich can be viewed as a Spacetime Construction package, 
to parts 1)-5) being a Configurational Relationalism package and part 4b) being a resolution of Temporal Relationalism.

\section{Conclusion}\label{Conclusion}

Temporal Relationalism (TR) compatibility of the other local classical Problem of Time facets has been completed.  
This was already covered in \cite{FEPI, TRiPoD, AM13} for the Configurational Relationalism (generalization of Thin Sandwich), Constraint Closure, Problem of Observables or Beables 
and Spacetime Construction facets.
[The Problem of Beables has been rendered TR compatible, not solved, in the case of GR.]
The current Article has now addressed TR compatibility of foliation issues also. 
This forms the ordering of facets in Fig \ref{Gates-2}.e).  

\mbox{ }

\ni In this approach it makes sense to use not ADM's split             with its lapse and shift 
but rather the TR implementing (TRi)             A   split \cite{FEPI} with its cyclic differentials of the instant and of the frame.
In the current Article, I recast the theory of foliations in TRi form.  
Here the time flow vector, also conceived of as a deformation vector, is the primary TRi homothetic covector from which the cyclic differential replacements for the lapse and shift follow.
Refoliation Invariance, bubble time and the Hojman--Kucha\v{r}--Teitelboim approach to GR from deformation first principles carry over to TRi form.  
The Machian Thin Sandwich can be extended to to give an emergent TRi version of Kucha\v{r}'s universal hypersurface kinematics. 
Here, Kucha\v{r}'s hypersurface derivative is reinterpreted as Barbour's Best Matched derivative by hypersurface tensor duality.
Tilt is reinterpreted as spatial gradient of the change of the instant. 
Finally, the extrinsic curvature of derivative coupling is reinterpreted as a comparison of each geometrodynamical change with the sufficient totality of locally relevant change 
that the emergent Machian generalized local ephemeris time is abstracted from.

\mbox{ }

\ni Natural successors of this Article are as follows.

\mbox{ } 

\ni A) TRiCQM (canonical quantum mechanics) \cite{ABook} e.g. along the lines of geometrical quantization \cite{I84}, 
and probably starting with the semiclassical case as regards further details.

\ni B) Rendering all steps global -- or at least detailing which steps fail to be global for what reasons --
and repeating the process assuming less levels of mathematical structure are also frontiers.
The last of these is vast \cite{ASoS, AGates, AMech}, and can be taken to be the offspring of Isham's removal of layers of structure with the Kucha\v{r}--Isham \cite{Kuchar92, I93} 
characterization of the Problem of Time facets (updated in \cite{APoT2, APoT3, ABook}).
Ie what form do these facets, and underlying Background Independence notions, take at each level of mathematical structure? 
Do they persist to all levels of mathematical structure, or, if not, at which levels do they shed significant notion of time properties or further exit the realms of technical tractability?
In particular, upon realizing that Problem of Time comes from incorporating Background Independence into Physics being conceptually and technically difficult.
Thus it is clear that the matter of Background Independence can be posed to all levels of mathematical structure and that it is very interesting and foundational to do so...

\mbox{ } 

\ni The current Article completes the relational approach to the Problem of Time at the classical level, modulo Dirac beables construction and global caveats.
To commemorate the 30th anniversary of \cite{IK85}, I end by summarizing this as follows.
{\it Cum grano salis}   \cite{ABeables, AGlob, ASoS} 
{\it finis coronat opus}: \cite{BSW, BB82, B94I, RWR, APoT2, AConfig, AM13, AGates, APoT3, ABeables, ABeables2, TRiPoD} and the present Article.

\mbox{ } 

\ni{\bf Acknowledgements} To those close to me gave me the spirit to do this.  
And with thanks to those who hosted me and paid for the visits: Jeremy Butterfield, John Barrow and the Foundational Questions Institute.
Thanks also to Julian Barbour, Jeremy Butterfield, Sean Gryb, Chris Isham and Flavio Mercati for a number of useful discussions over the years.  


\end{document}